\documentclass[sigconf]{acmart}

\AtBeginDocument{%
  \providecommand\BibTeX{{%
    \normalfont B\kern-0.5em{\scshape i\kern-0.25em b}\kern-0.8em\TeX}}}

\setcopyright{acmcopyright}
\copyrightyear{2023}
\acmYear{2023}
\setcopyright{acmlicensed}\acmConference[UIST '23]{The 36th Annual ACM
Symposium on User Interface Software and Technology}{October 29-November
1, 2023}{San Francisco, CA, USA}
\acmBooktitle{The 36th Annual ACM Symposium on User Interface Software
and Technology (UIST '23), October 29-November 1, 2023, San Francisco, CA,
USA}
\acmPrice{15.00}
\acmDOI{10.1145/3586183.3606772}
\acmISBN{979-8-4007-0132-0/23/10}

\usepackage[capitalize, noabbrev]{cleveref}
\usepackage{tabularx}
\usepackage{booktabs}
\usepackage{multirow}
\usepackage{tikz}
\usepackage{caption}
\usepackage{subcaption}

\usepackage{graphicx} %
\usepackage[skip=3pt]{caption}
\usetikzlibrary{shadows}
\begin{document}

\definecolor{changeNoteColor}{rgb}{0.1,0.6,1}
\newcommand{\changenote}[1]{#1} %

\renewcommand{\shortauthors}{Dang et al.}
\newcommand{\formativeN}{4}
\newcommand{\mainN}{13}
\newcommand{\system}{WorldSmith}
\newcommand{\globalview}{\textit{Global Tile View}}
\newcommand{\detailview}{\textit{Detail Editor View}}
\newcommand{\RQ}[1]{\textit{$({RQ#1})$}}

\newcommand{\participant}[2]{$P_{#1}$}
\newcommand{\pquote}[3]{\textit{``#1''} (\participant{#2}{#3})}

\newcommand{\designgoal}[1]{\textbf{\textit{D#1}}}

\definecolor{cinterface}{HTML}{000000}

\title[\system{}]{\system{}: Iterative and Expressive Prompting for World Building with a Generative AI}

\author{Hai Dang}
\email{hai.dang@uni-bayreuth.de}
\orcid{0000-0003-3617-5657}
\affiliation{%
  \institution{University of Bayreuth \& Autodesk Research}
  \city{Bayreuth}
  \state{Bavaria}
  \country{Germany}
}

\author{Frederik Brudy}
\email{frederik.brudy@autodesk.com}
\orcid{0000-0002-3868-0967}

\affiliation{%
  \institution{Autodesk Research}
  \city{Toronto}
  \state{Ontario}
  \country{Canada}
}

\author{George Fitzmaurice}
\email{george.fitzmaurice@autodesk.com}
\orcid{https://orcid.org/0000-0002-2834-7757}

\affiliation{%
  \institution{Autodesk Research}
  \city{Toronto}
  \state{Ontario}
  \country{Canada}
}

\author{Fraser Anderson}
\email{fraser.anderson@autodesk.com}
\orcid{https://orcid.org/0000-0003-3486-8943}

\affiliation{%
  \institution{Autodesk Research}
  \city{Toronto}
  \state{Ontario}
  \country{Canada}
}

\begin{abstract}
  Crafting a rich and unique environment is crucial for fictional world-building, but can be difficult to achieve since illustrating a world from scratch requires time and significant skill. We investigate the use of recent multi-modal image generation systems to enable users iteratively visualize and modify elements of their fictional world using a combination of text input, sketching, and region-based filling. \system{} enables novice world builders to quickly visualize a fictional world with layered edits and hierarchical compositions. Through a formative study (\formativeN{} participants) and first-use study (\mainN{} participants) we demonstrate that \system{} offers more expressive interactions with prompt-based models. With this work, we explore how creatives can be empowered to leverage prompt-based generative AI as a tool in their creative process, beyond current "click-once" prompting UI paradigms.
\end{abstract}

\ccsdesc[500]{Human-centered computing~Human computer interaction (HCI)}

\keywords{Multi-modal image generation, Fictional world-building, AI-assisted creativity}

\begin{teaserfigure}
  \centering
  \includegraphics[width=0.9\textwidth]{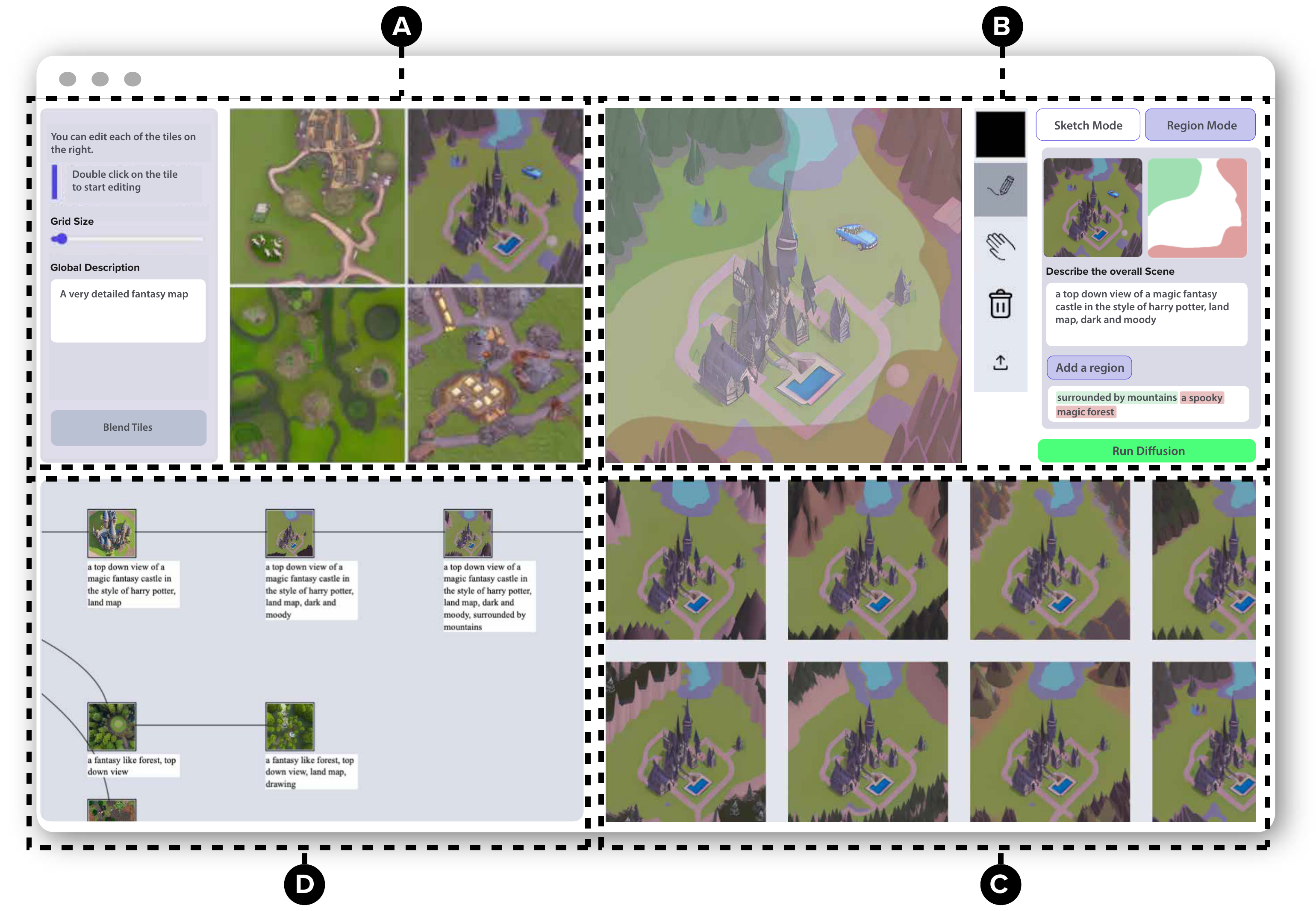}
  \caption{
  The workflow and high-level interface of \system{}. The user selects one of the four image tiles in the \textit{Global Tile View} (A), and iteratively edits this tile with text prompts, sketching, and region-painting tools available through the \textit{Detailed Tile Editor}  (B). All generated images are collected in the \textit{Results View} (C) which allows the users to re-use previous image assets. Furthermore, the \textit{Tree View} automatically captures each new image generation request (D). After creating all tiles the user returns to the \textit{Global Tile View} to blend all tiles into a single image.
  }
  \Description{A screenshot of WorldSmith depicting the high-level interface. The screenshot shows the Global Tile View, Detail View, Tree View and Result View.}
  \label{fig:teaser}
\end{teaserfigure}

\maketitle

\section{Introduction}

Fictional world-building is the process of constructing a fictional universe with its unique history, geography, culture, and rules \cite{Ekman2019VitruviusC}. 
In his seminal essay - \textit{On Fairy Stories} - Tolkien emphasizes the crucial role of an \textit{imagined place} for an engaging fantasy plot. This observation extends into the present, where interesting and intricate worlds are essential for the success of games, movies, and other forms of entertainment.

There are many reasons why people consume fiction with imaginary worlds. While some just enjoy the creative challenge of conceiving an interesting world itself, others appreciate the enormous vibrant community \cite{RodrguezFuentes2022WhyDP}. Forms of engagement include writing fan-fiction \cite{Samutina2016FanFA}, developing indie games \cite{Domnguez2019TheDO}, or running the popular table-top game Dungeon and Dragons (DnD)\cite{dndgygax1989}. 

Fictional worlds can be conveyed through various media, and creating visual artwork that depicts part of the imagined world is one of them. Visually materializing the world helps others find common ground when discussing and collaborating \cite{Kulkarni2023AWI}. However, illustrating these worlds is time-consuming and particularly difficult for novice world builders, who often lack the artistic skill or the experience to create coherent worlds. But even experienced world builders must invest significant time and effort into materializing their ideas.

With current illustration tools such as Adobe Photoshop or Adobe Illustrator, users often have to make many fine-grained edits to create their world. This is time-consuming and requires a high level of expertise. To relieve users from designing on the micro-level, many domain-specific tools have been developed, such as procedural terrain generation tools \cite{worldanvil2023, worldmachine2022, quadspinner2022gaea}, which algorithmically generate varied terrains based on a fixed set of rules or grammar \cite{Raffe2015IntegratedAT, Togelius2013ControllablePM}. Other tools have been developed to support the creation of 3D worlds \cite{Zhang2020FlowMaticAI, Wang2013DIYWB} and game levels \cite{Torrado2019BootstrappingCG}. However, these tools still operate procedurally, preventing users from defining their world using a high-level semantic description.

Recent image generation models such as Dall-E \cite{Ramesh2022HierarchicalTI},
Stable Diffusion \cite{rombach2021highresolution}, Imagen \cite{Saharia2022PhotorealisticTD}, and Midjourney \cite{midjourney2023} are now capable of generating high-quality images based on simple natural language text prompts to guide the image generation process semantically. Thus prompting has evolved as a new interaction paradigm with the advent of large pre-trained text \cite{RoberGPT3WGF2021} and image generation models. 

Prompt-based image generation models have become increasingly popular, but the prevailing interaction paradigm with these models is limited to a "click-once" text prompt interface. This approach assumes that users can provide a complete and accurate description of the desired visual imagery upfront. However, world-building is an iterative process, and this simplistic approach may not be adequate \cite{zaidi2019worldbuilding}. More expressive techniques are needed to interact with these models to address this challenge. One promising approach is incorporating additional input modalities, such as sketching or other graphical interfaces, to allow users to convey their design. However, the impact of such multi-modal systems on the world-building process and the behavior of users when defining prompts for generative AI  models remains an open question.

To this end, we designed and built \system{} (\cref{fig:teaser}), a tool to support world-builders \changenote{generate an image of their envisioned fictional worlds through multi-modal inputs, including text input, sketching, and region painting}. To accommodate their iterative and piece-by-piece workflow, \system{} was designed to reinforce two key concepts 1) hierarchical generation of multiple image tiles and 2) layered editing of individual image tiles. To evaluate the utility and to observe users' prompting behavior with \system{}, we conducted a first-use study with \mainN{} participants.

In summary, we contribute the following:
\begin{itemize}
    \item \system{}, a multi-modal tool that enables users to iteratively design and refine complex fictional worlds using layered editing and hierarchical compositions with a prompt-based model that uses text, sketches, and region masks as inputs.
    \item Insights from a formative and first-use study, demonstrating how \system{} facilitates interactive prompting with text input \changenote{and additionally with non-textual interaction such as sketching and region painting, to disambiguate text prompts for generative AI.}
\end{itemize}

\section{Related Work}
This work draws upon the domains of world-building, scene generation and human-AI co-creativity.

\subsection{Image Synthesis Techniques}

To facilitate working with multiple generated images, \system{} uses various image composition techniques, including inpainting, which involves filling in missing or damaged areas of an image by generating plausible content based on the surrounding context \cite{Yu2018GenerativeII}. Inpainting has been applied to automatically colorize rough sketches \cite{Sangkloy2016ScribblerCD} and remove objects from photographs \cite{Criminisi2003ObjectRB}. Another related technique is outpainting, which generates new content beyond the boundaries of an image \cite{Ying2020180degreeOF}. Pre-trained image models have also been investigated for their ability to perform visual conceptual blending \cite{Ge2021VisualCB}, which involves blending visual concepts to generate new content, such as an "amphibious vehicle" resulting from blending "a boat" and "a bus." Other research has explored blending for creating symbols \cite{Chilton2019VisiBlendsAF} and for sketches \cite{Karimi2019ACM, Cunha2017APA}. Multiple inputs, including fully segmented images with corresponding annotations \cite{Gafni2022MakeASceneST, balaji2023ediffi}, can be used to synthesize images. Motivated by the world-building workflow, we investigate how various image synthesis techniques work together to support users in their world-building process. Specifically, we blend multiple image tiles to create a larger composition, generating new content beyond the boundaries of individual images while staying within the boundaries of the target set as a whole.

\subsection{Prompting Pretrained Generative Models}
Recent developments in natural language processing have shown that Pre-trained Large Language Models (LLMs) can solve multiple tasks without the need for specific training for each task. This can be achieved by using text prompts in natural language, as demonstrated by \citet{Brown2020gpt3}. 
Generating effective text prompts is a challenging task, not just for generating text \cite{liu_pre_train_2021}, but also for generating images. Although the internet community has developed several prompting strategies to create more targeted images, such as incorporating resolution-related terms like \textit{4k} or \textit{Unreal Engine}, recent research has proposed techniques to automatically refine these prompts through prompt engineering \cite{jiang_how_2020, yuan_bartscore_2021, liu_what_2021, haviv_bertese_2021} or interactive methods \cite{liu2021design, liu2022opal}.
While, many interactive prompt-based tools only support uni-modal input \cite{Coyne2001WordsEyeAA, Sharma2018ChatPainterIT, liu2022opal}, two recent surveys \cite {Kulkarni2023AWI, Ko2022LargescaleTG} independently called for also exploring multi-modal affordances of prompt-based models. To this end, \citet{liu20223dalle} designed 3DALL-E to support image prompts in addition to text inputs, by taking snapshots of model-objects workspace in their workspace and generating variations of that input. \citet{Zhang2022StoryDrawer} introduced StoryDrawer a co-creative drawing system to support children in creative storytelling through interacting with an AI through a conversational dialog and drawing. In the current literature, investigations towards more expressive prompting are scarce. However, enabling users to better express themselves when interacting with prompt-based models is crucial to support more complex workflows such as world building. Therefore, we add two new dimensions of expressive prompting to the current literature, namely hierarchical prompting and spatial prompting.

\subsection{Scene Generation}
Although our focus is on the visual representation of 2D fictional worlds, previous research has already examined text-based worlds \cite{Jansen2021ASS, Ct2018TextWorldAL} and virtual 3D worlds \cite{Zhang2020FlowMaticAI}. Anticipating the emergence of language-based 3D scene generation systems, \citet{Coyne2001WordsEyeAA} presented \textit{WordsEye}, a tool that automatically converts text into 3D scenes. However, this tool relied on a vast database of pre-existing 3D models and poses. In contrast, fictional world-building typically involves the creation of new artwork. Consequently, this method may face limitations when dealing with unstructured 2D fictional worlds or unconventional layouts. According to a recent survey \cite{Zakraoui2019TexttopictureTS}, interactive text-to-scene systems are relatively scarce, while most related work has concentrated on automated approaches \cite{Joshi2006TheSP, Gupta2018ImagineTS, Lee2013SketchStoryTM}, neglecting the role of humans in the creative process. Nonetheless, there are some examples of systems that adopt a more human-centric perspective \cite{Qiao2019LearnIA, Coyne2001WordsEyeAA}. Our system, \system{}, is intended to assist users in creating rather than substituting for them.

There has been growing interest in the development of interactive scene-generation systems. One approach is to use a scene graph to generate images, as explored in \cite{Mittal2019InteractiveIG}. Another interesting direction in interactive image generation is the use of chat interfaces \cite{Sharma2018ChatPainterIT, ElNouby2018KeepDI}. However, fictional worlds often have a spatial component, but it has been found that human language for expressing spatial relations is often ambiguous and subjective \cite{Patki2019LanguageguidedSM, Liu2022ThingsNW}. We designed \system{} to allow users to draw their spatial knowledge in addition to text input.

\subsection{Human-AI Co-Creativity}
Co-creative systems that involve both humans and Artificial Intelligence (AI) entail collaboration where each party contributes their capabilities to the creative process. AI systems can assist with generating ideas \cite{elephant_singh, Yuan2022WordcraftSW}, provide inspiration \cite{Gero2021SparksIF, Yang2022AIAA, Jeon2021FashionQAA}, and support internal reflection on the content \cite{Liu2021IncorporatingAR, Dang2022BeyondTG}, while humans can provide subjective judgment and critical thinking. Recently developed co-creative tools include FashionQ \cite{Jeon2021FashionQAA} for ideation in fashion design, and WeToon \cite{Ko2022WetoonAC} which enables users to generate sketches through direct manipulation of a graphical user interface. Other tools employ prompt-based AI models to support users in their design process \cite{liu20223dalle, liu2022opal}. However, \citet{Jakesch2023CoWritingWO} found that generative model biases can influence users' behavior and lead them to choose the most convenient option, often the first generated content item. This insight further highlights the need for the careful evaluation and design of co-creative systems. To this end, various frameworks have been developed to support the design of a creative partner \cite{Rezwana2022DesigningCA, Llano2022ExplainableCC, Amershi2019GuidelinesFH}, given the challenges of developing effective AI. Moreover, several works have suggested to also logging users' interactions \cite{Lee2022CoAuthorDA, Dang2022GANSliderHU, Dang2023ChoiceOC, liu20223dalle} to evaluate their behavior with co-creative AI systems. 
With \system{}, we contribute a creativity tool to support users' world-building workflow through iterative and expressive interactions.
\section{Formative Study}\label{section:formative_study}
A formative study was carried out to gain insight into the approaches and perceptions creators have in constructing and defining their fictional worlds.

\subsection{Participants and Procedure}
A total of N=\formativeN{} individuals participated in remote interviews, with ages between 25-52 years. Participants were recruited through internal email lists and one external professional was recommended by personal contacts. All participants reported prior experience in world-building, either as a hobby or within a professional context designing landscapes or animating game assets. Two participants spent less than an hour per week on world-building tasks, one spent 1-5 hours per week, and one spent at least 5 hours per week constructing worlds.

During the study, semi-structured interviews were conducted to investigate how participants plan their world-building process. On average, each interview took 45 minutes. 
The guiding questions were focused on how the study participants characterized world-building and the methods they employed throughout their world-building process.

\subsection{World Building Process}\label{section:formative_results}
From the results of the interviews, we found that participants had different motivations for building worlds as well as different tools used, though many followed similar processes. While some participants (\participant{1}{}, \participant{3}{}) found joy in developing interesting and fun worlds to share with their friends, other participants created intricate worlds as part of their profession. For example, \participant{2}{} taught about a landscape design class at a university and builds terrain maps to investigate how architectural structures evolve over time based on the surrounding environment. \participant{4}{} is a professional game animator who has animated multiple 3D fictional worlds.

\subsubsection{Finding Inspiration and Re-using Assets}
When it comes to finding inspiration for their creative work, many commented that they often turn to the internet. For example, they reported using search engines or browsing related blogs and online communities like Reddit in search of visual imagery that sparks their imagination (\participant{1}{}, \participant{2}, \participant{3}{}). In a Dungeon and Dragons campaign, \participant{1}{} had the primary responsibility of constructing the game's world. However, they found creating visual assets tedious and time-consuming. As a result, \participant{1}{} frequently resorted to using pre-existing assets found online and noted that there may be copyright issues with re-using these assets.

\subsubsection{Refining Ideas}
The initial image or inspiration is often vague and often requires many iterations before it becomes something concrete. From their initial spark, participants began asking themselves questions about the fictional world they wanted to build, using an image sketch or list of notes as a starting point to generate further ideas and details. As one participant explained, \pquote{Sometimes a piece of visual artwork strikes me and I find myself asking: What is happening in that image? What is the character doing in that scene?}{4}{} This iterative process of questioning and building upon ideas aids the development of richer and more detailed fictional worlds.

\subsubsection{Inductive Rather Than Divergent World-Building Process}
On a macro level, rather than exploring many divergent creations, it is more common to fix key characteristics such as the time and age, or style of the world (fantasy world, science fiction world) once at the beginning of the world building process. Fine details of the world such as individual fauna, flora, and their spatial composition are subject to more frequent changes. However, the further a creator is in the design process, i.e. the more complex the fictional world is, the less likely are major changes in the image composition, because such changes would introduce too much re-work.

\subsubsection{Throwaway Prototypes}
After the initial idea finding phase, a key activity is creating disposable prototypes quickly. There are many tools to support world builders to accomplish their task, but these tools are highly specialized and often have a steep learning curve to master them fully. Participants in our study reported that they use multiple tools during their world building process. Nevertheless, all participants often started with a rough outline of the world they want to build using pen and paper only, because it allows them to quickly note and depict ideas. These outlines may only include a list of notes (\participant{3}{}), but more often also include graphical elements such as a mind map (\participant{4}{}) or layout sketch with text annotations that describe which elements should appear in their artwork later on (\participant{1}{}, \participant{2}{}).

\subsubsection{High-Fidelity Artwork}
After an initial stage of ideation, the desired level of professionalism and the complexity of the fictional world determines which tools they use. As a hobby Dungeon and Dragons game master, \participant{1}{} use specialized online world-building tools for generating graphical game elements such as maps \cite{inkarnate2023, worldanvil2023}, assets \cite{dungeonfog2023}, and characters \cite{dndbeyond2023}. Furthermore, \participant{1}{} felt overwhelmed with learning all the tools required to create various elements for the DnD campaign. \participant{3}{} is an experienced software developer and built a computational agent that procedurally created a DnD map. During the DnD campaign \participant{3}{} would use a printed version of the previously generated map and let players spontaneously draw additional game elements on it. Although it saves time, \participant{3}{} wanted to have an image for the player to set the "mood" for the current DnD campaign. For landscape design, \participant{2}{} use a range of terrain editing software \cite{worldmachine2022, terragen2022, quadspinner2022gaea, bitethebytes2023WC} to model photo-realistic landscapes. Here, \participant{2}{} noted that, although these tools produce highly realistic 3D terrains, this comes at the cost of a high learning curve.

\section{Design Goals}\label{section:design_goals}
We formulated the following design objectives, based on the processes identified during our formative user study (\cref{section:formative_results}), to assist users in constructing their fictional worlds.

\textbf{D1 - Support Multi-Modal Input}
World-building includes multiple steps that require prototypes of different fidelity. Early prototypes are usually coarse and are mainly text driven, sometimes also including simple sketches. Our system needs to support multiple input modes to allow users to express their design intent.

\textbf{D2 - Supporting Iterative Refinement}
The system needs to allow users to continually incorporate new details into their world. Therefore, the prototype should facilitate the ability to make layered revisions to images that users have previously created.

\textbf{D3 - Support Visual Asset Generation}
In order to facilitate the visualization of intricate worlds, the prototype should empower users to create new visual assets to populate their world. 

\textbf{D4- Enable Hierarchical Composition}
The formative study indicated that world builders typically work on various levels of detail. At the macro level, they establish the general layout of the world, while at the micro level, they determine the specific components present within the world. Hence, our prototype should allow users to engage hierarchically in the design process.
\section{\system{}}\label{section:concept}
We designed our prototype (\cref{fig:teaser}) to support users' world-building workflow by allowing them to focus on different sub-components of their world and perform layered edits. Through multiple generated images users iteratively refine their initially vague ideas. An interactive \textit{Tree View} allows them to introspect their past actions and branch out to create new visual assets all in one application.

\begin{figure}
  \includegraphics[width=\linewidth]{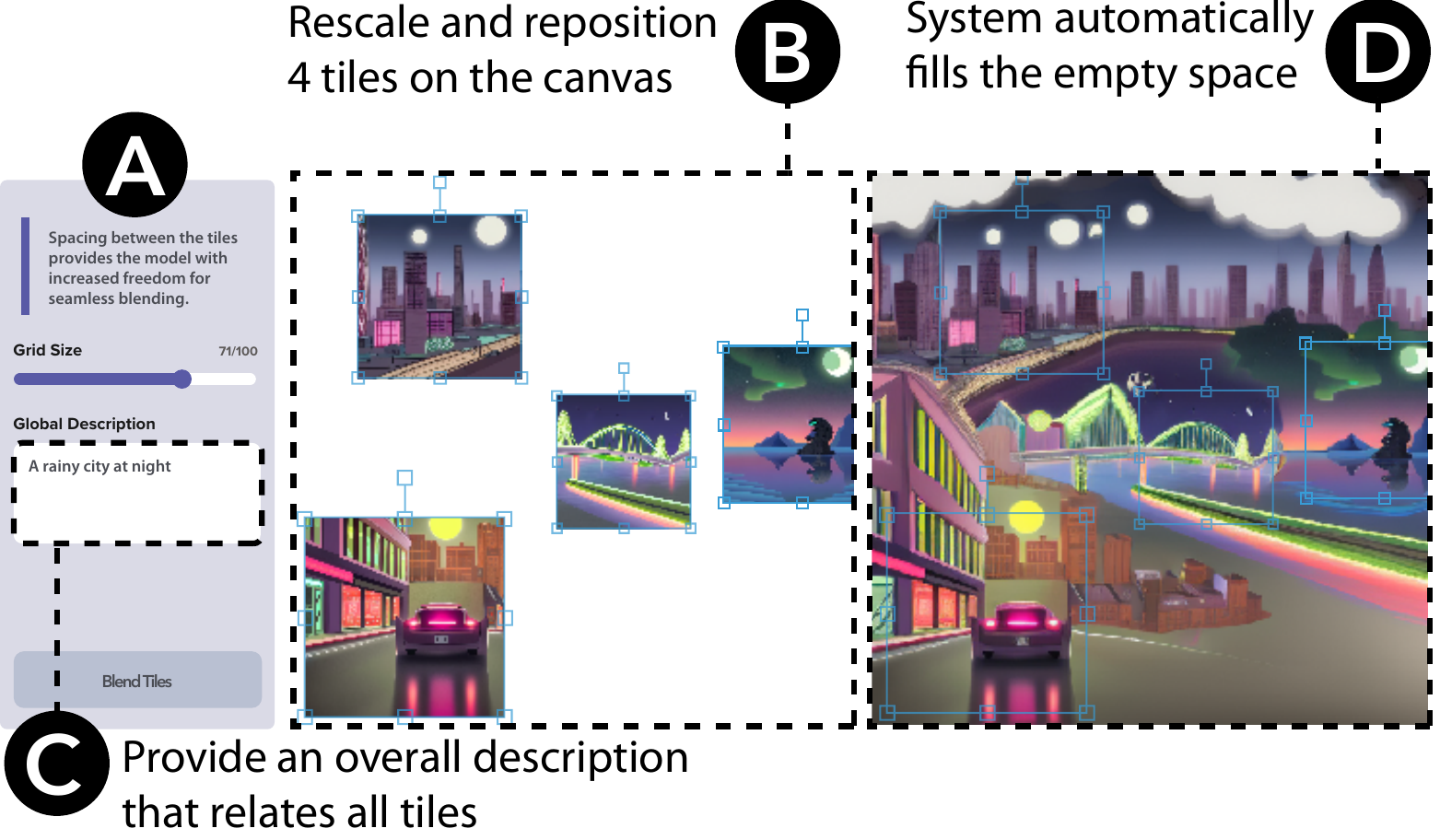}
  \caption{Figure depicts the \textit{Global Tile View}, where all the tiles have already been created by \participant{11} (B). The tool panel on the
left-hand side facilitates the control over space between individual tiles (A) and, further, allows for a description of how the tiles 
should be blended together (C). The text prompt reads: ``a rainy city at night.''}
  \Description{A screenshot showing four image tiles. Tiles are blended into a cohesive image.}
  \label{fig:global_tile_view}
  \vspace{-1em}
\end{figure}

\subsection{Global Tile View}\label{section:global_tile_view}

Inspired by the game Carcassone, \system{} lets users create a world image with multiple \textit{Image Tiles} (\textbf{D4}, \cref{fig:global_tile_view} (B)). \system{} includes four image tiles \changenote{which gave participant in our user study (\cref{section:main_study}) enough time to work on each tile in detail. However, our concept also allows for more image tiles.} Tiles are initially aligned in a grid but can be resized and moved on the canvas.

\system{} supports multi-level editing with the \globalview{} which allows users to blend multiple tiles together to form a cohesive world (\textbf{D4}). We decided to separate the image tile composition from the image tile creation, as they conceptually represent different layers of abstractions. In addition, this division allows users to concentrate on broader objectives, such as combining various tiles that depict the creative vision of the world, while deferring the intricate editing process for each image tile. Typically, these components are guided by a narrative framework; for instance, a user may wish to construct a map featuring a forest, a lake, and mountains. Once blending is complete, the result is shown next to the global tile view, and users can return to individual tiles for further editing if necessary.

\textit{Blending Tiles} Users can blend tiles by providing a text prompt (\cref{fig:global_tile_view}) and adjusting the empty space between them. The system fills the space between tiles based on the created tiles and a text prompt.  

\textit{Resizing and Repositioning 
Tiles}
The \textit{Grid Size} slider controls the amount of empty space, with more space providing more blending space. Users can also resize and reposition tiles on the canvas for added flexibility.

\subsection{Detail Tile Editor}\label{section:detail_view}
The \detailview (\cref{fig:teaser}B) allows users to focus on a specific part of the world within a larger composition. It provides several tools that allow users to generate content using text, sketching, and masking (\textbf{D1}).

\subsubsection{Text Prompt Editor}\label{sction:prompt_editor}

The text editor of \system{} comprises of two sections: a global scene description area and a specific region description area for users to provide more focused descriptions.

\textit{Overall Scene Description}
The overall scene description text box provides a plain text entry box for users to generate image content quickly. This is a common UI pattern already found in many recent generative image generation tools \cite{Ramesh2022HierarchicalTI, midjourney2023}.

\textit{Region Description}
The region description permits users to spatially specify where content should generate on the canvas. Adding a region inserts an empty text segment with a newly assigned color (see \cref{fig:teaser}B). Users can modify this segment by typing a description and only affecting the outlined area. Regions can be drawn to visually link to the corresponding text segment color.

\subsubsection{Large Canvas}
\begin{figure}
    \centering
    \includegraphics[width=\linewidth]{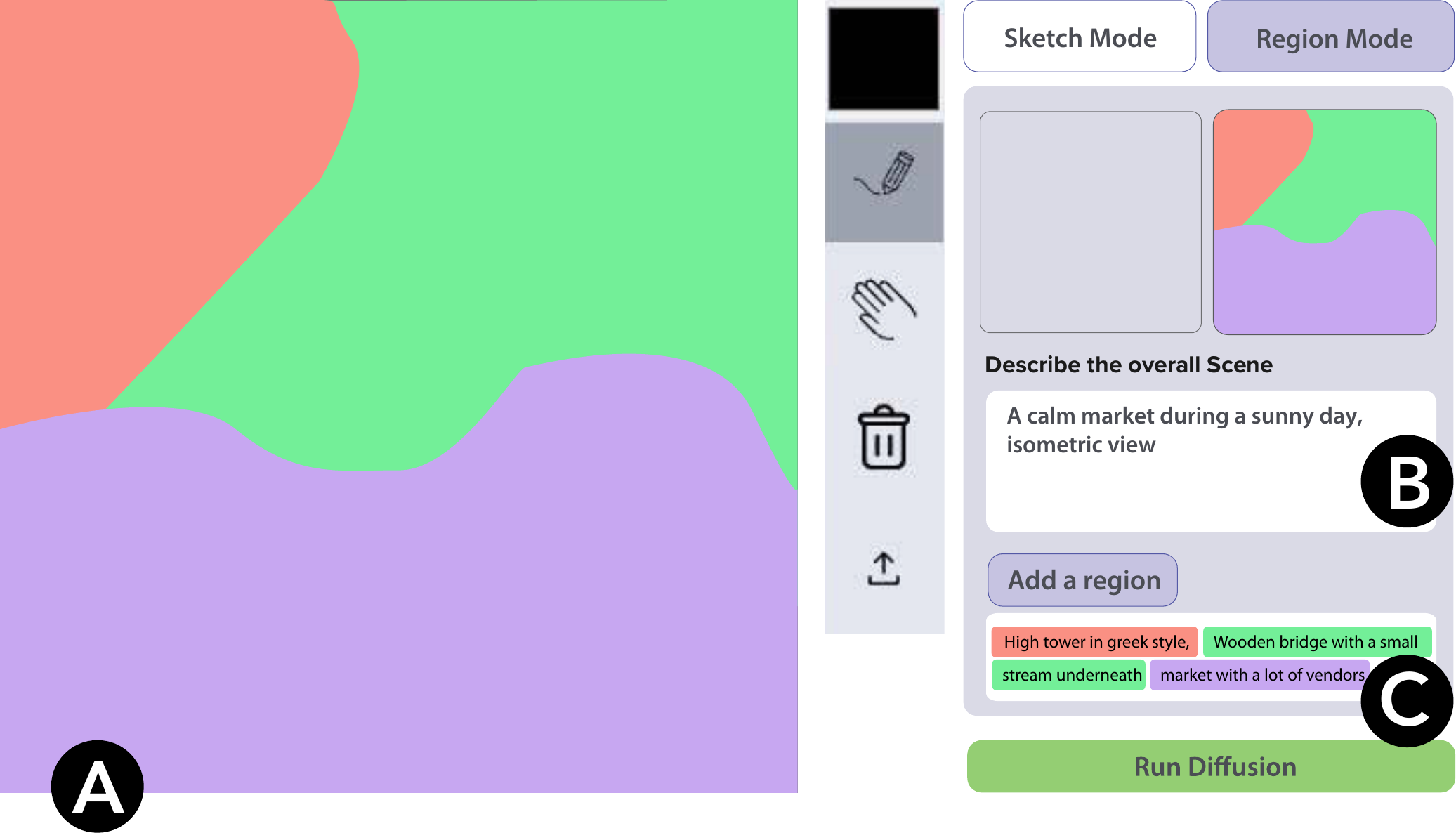}
    \caption{An example of a region segmentation (A) from $P_{6}$ with the corresponding scene (B) and region (C) description in the prompt editor. The regions and corresponding text segments are shown with the same color.}
    \Description{A screenshot showing 3 user-created regions highlighted in different colors. For each region, the user has created a corresponding region description.}
    \label{fig:region_drawing}
    \vspace{-0.5em}
\end{figure}

A large multi-purpose canvas allows users to draw sketches and iterate on their generated images (\cref{fig:region_drawing}). Users can choose between a sketch mode or a region mode when iterating on an image (\textbf{D2}). Both modes support a textual description (see \cref{section:detail_view}) that instructs the system how to interpret users' spatial inputs.

\begin{figure}
    \centering
    \includegraphics[width=\linewidth]{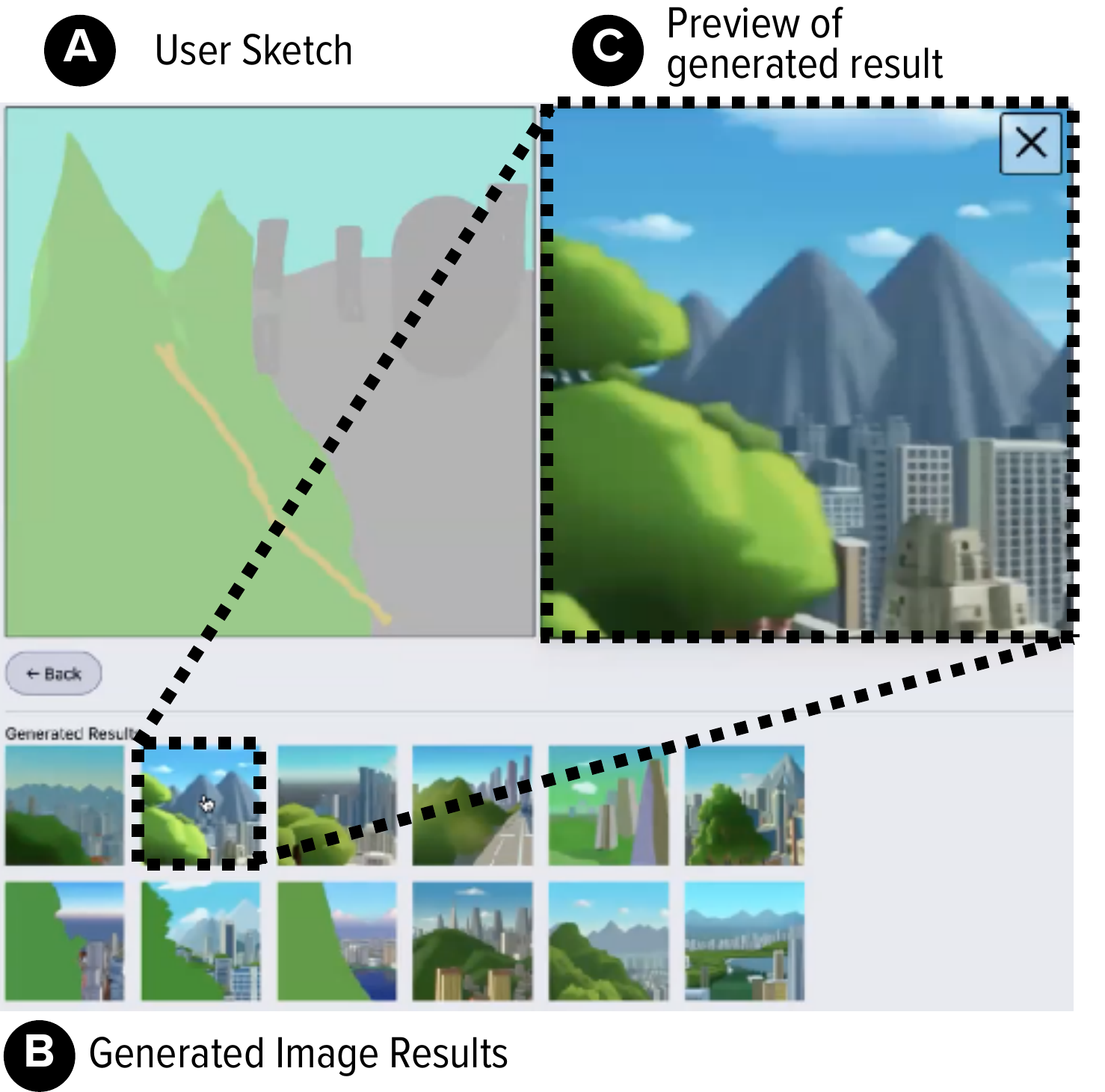}
    \caption{(A) An example of a sketch from $P_{3}$ with the overall scene text prompt: ``A skyline view of a city in the Caribbean, it has a chain of mountains in the left, and next to them there is a big city, it has a lot of skyscrapers, a path is coming down from the mountain and integrating into the city, Anime style''. (B) The generated results are shown at the bottom and (C) enlarged when hovered over.}
    \Description{A screenshot showing a user sketch and the resulting generated images based on the sketch.}
    \label{fig:sketch}
    \vspace{-1em}
\end{figure}

\textit{Sketch Mode}
The sketch mode lets users draw sketches using a pen tool (see \cref{fig:sketch} c). \system{} takes the sketch input and a textual description to generate images that are similar to the user's drawn images but adds more detail to the generated image (\cref{fig:sketch}). This allows users to coarsely sketch their image while the system generates a higher fidelity image. They can also drag existing images and generated images into the sketch canvas to create variations of that image to quickly explore alternative image generations.

\textit{Region Mode}
In the region mode users can draw a region on the canvas by using one of several region brushes: 1) The \textit{Pencil Brush} allows users to draw simple strokes. Each stroke corresponds to one of the text segments previously defined when users created a new region (see \cref{fig:teaser}B), 2) The \textit{Hull Brush}  computes the convex hull of all brush strokes combined since the hull brush was selected. This brush allows users to quickly select large regions. 3) The \textit{Lasso Brush} creates a closed path and lets users quickly draw closed shapes. The masks are only visible when the region mode is active (see \cref{fig:region_drawing})

\subsubsection{Generating Images}
After users are done defining their inputs which may include a scene description, a sketch, as well as a few region descriptions, they can generate an image by clicking on the \textit{Run Diffusion} button (\cref{fig:teaser}B). 

\subsubsection{Results View}
The \textit{Results View} is a collection of all images that users have generated displayed as small thumbnails in a scrollable grid. Hovering over a thumbnail displays a larger preview of the image next to the canvas, allowing users to directly compare the differences between the two. When new images are fetched, the \textit{Results View} temporarily greys out to indicate that new content is currently being generated. Once generated, these images will display in the \textit{Results View} so users can reuse it in all their edits.

To insert a selected image into the canvas, users can double-click on the corresponding thumbnail or simply drag the image onto the canvas.

\begin{figure}
    \centering
    \includegraphics[width=\linewidth]{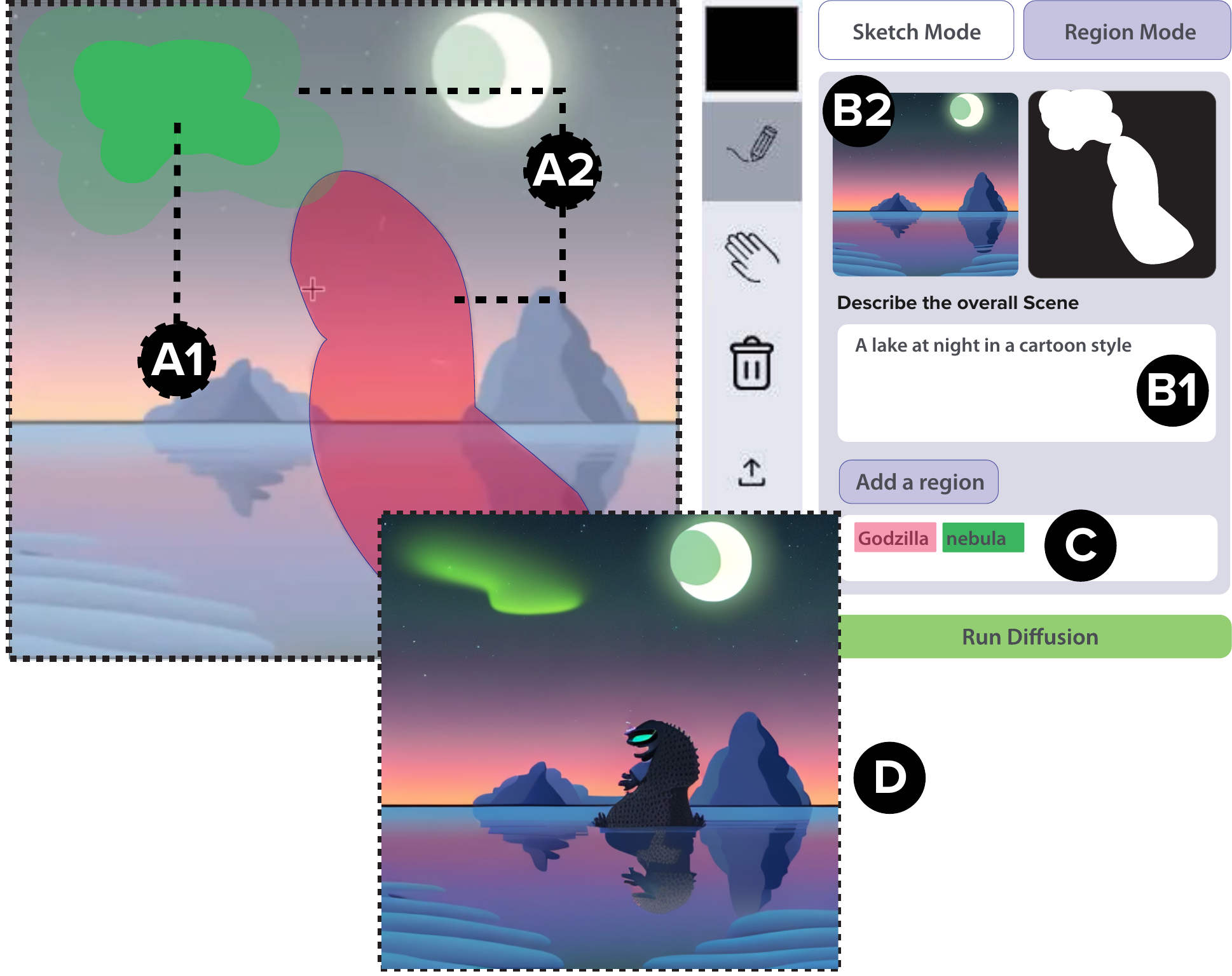}
    \caption{An example of adding sketch information to refine a generated image (B2). $P_{11}$ used a green pen to sketch on the image (A1) and a region tool to highlight parts of the image (A2), thereby providing  extra information to guide the generation of a green nebula and Godzilla. The scene description reads: ``a lake at night in cartoon style'' (B1) and the regions include \textit{Godzilla} and a \textit{nebula} (C). (D) The system redrew the user's input, adding a nebula and Godzilla that matches the style of the image.}
    \Description{A screenshot collage showing the layered edits a user has made on an image.}
    \label{fig:godzilla}
    \vspace{-1em}
\end{figure}

\begin{figure*}
    \centering
    \includegraphics[width=\textwidth]{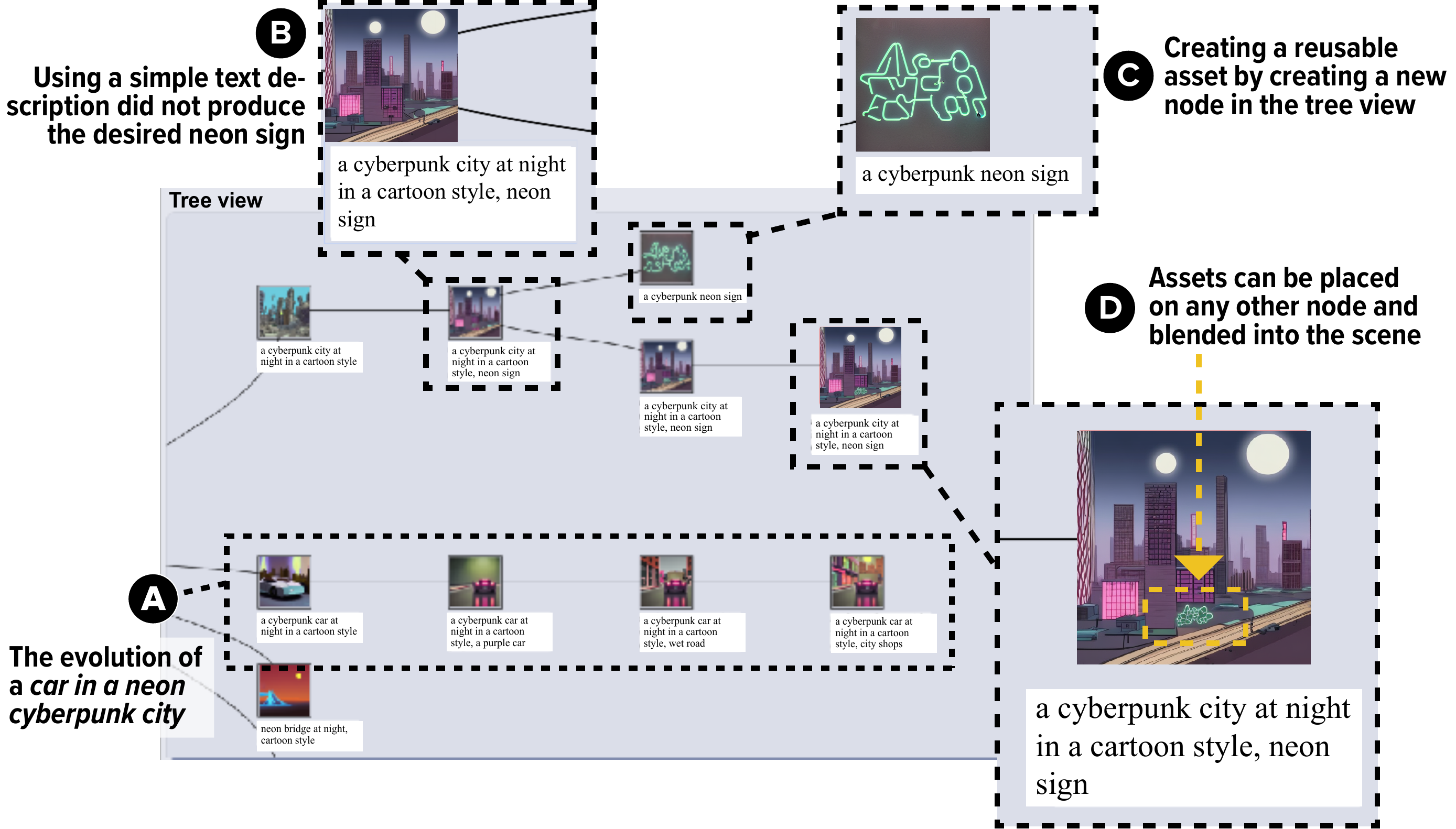}
    \caption{(A) The tree view shows all user inputs and interactions in a tree diagram. This allows users to retrace their creation process, branch out, and create alternative versions of any tile or entirely new assets. 
    (B) During their creation of a cyberpunk city, $P_{11}$ found the neon sign to be lacking. They used the tree view feature to create a neon sign (C) and merged that newly created asset into the previously generated image scene (D).}
    \Description{A screenshot with enlarged elements that depict the contents of a tree view at different iteration stages.}
    \label{fig:treeview}
    \vspace{-1em}
\end{figure*}

\subsection{Tree View}
The tree view records all user actions for each tile (see \cref{fig:treeview}). Each node in the tree view includes an image preview and a detailed text description that contains both the scene and region descriptions, representing a snapshot of a tile.

The tree view supports reflective thinking by displaying previous inputs and allowing users to explore the evolution of their design throughout the session \cite{Burg2013InteractiveRF, Grossman2010ChronicleCE}. Moreover, it encourages divergent thinking by enabling users to create alternate iterations (\textbf{D2}) and explore new visual assets (\textbf{D3}). In the realm of text-based world simulation systems, knowledge graphs have been employed to represent the state of the world by mining existing storylines \citet{Ammanabrolu2020BringingSA}. \system{} automatically creates such a state tree and additionally enables users to manually extend this state tree.

Our description below outlines the user interactions that facilitated exploring and refining image tiles with the tree view.

\textit{Automatic node insertion}\label{section:add_node}
Whenever users execute the diffusion process, the system automatically updates the tree view. If the inputs to the system have been altered since the last image generation, a new node will be added to the currently selected tree node. This automated insertion upon update enables users to reference their previous interactions.

Furthermore, users have the option to manually insert nodes by selecting a node and clicking on the \textit{Add Node} button. This interaction pattern is useful for experimenting with new-generation ideas while preserving the previous generation. When inserting a new node, users can choose to iterate on an exact copy of the previous inputs, or they can start from scratch to create new visual assets and then blend it into their previously generated world (see \cref{fig:treeview}).

\textit{Selecting a Node}.
Instead of adding a new node, users can also double-click on an existing node in the tree view to load the inputs linked to that particular node into the detail editor view, enabling them to continue iterating on it (\textbf{D2}).

\textit{Pan and Zoom} To accommodate the expanding size of the tree view, we added a pan and zoom function to allow users to zoom in on specific nodes and examine their input. Users can zoom into the canvas using the mouse wheel and pan by clicking and dragging the canvas while holding the left mouse button. Upon generating a new node (see \cref{section:add_node}), the system automatically centers the viewport to present the node at the center of the screen, aiding users in maintaining focus and navigating the tree view more efficiently.

\textit{Preview generated images}
\system{} automatically displays a preview of all images generated for a particular node and its inputs when users hover over an image. This feature allows users to refer to the text prompt beneath the node while also previewing the content they have created.

\subsection{Usage Scenario}\label{sec:usage_scenario}
The following describes one possible scenario of how \system{} can be used. This scenario \changenote{compactly describes how \participant{11}{} in our user study (\cref{section:main_study}) worked with \system{}. It demonstrates how} the different components of \system{} interplay to enable a user to iteratively generate a visual representation of a fictional world, by using text based prompting, iterative workflows, and detailed editing using sketching and regions.

\participant{11}{}, an avid reader of fictional novels and a creative writer, struggles to convey the mood of her imaginative world to her collaborators. To illustrate her world and build a shared understanding with them, she uses {\normalfont \system{}}, to iteratively create a visual representation (e.g., a map view) of her world from scratch.

To start, she selects one of the four empty image tiles in the {\normalfont Global Tile View} 
(\cref{fig:teaser}~A) and enters ``\textit{a lake at night in a cartoon style}'' as the scene description (\cref{fig:godzilla}~B1). After generating and selecting her favorite initial scene it appears as an input preview (\cref{fig:godzilla}~B2). However, it is missing the all-important Godzilla. She switches to the {\normalfont Region Mode} (\cref{fig:godzilla}~A2, red) and enters the text description ``godzilla'' as a region descriptor (\cref{fig:godzilla}~C). 
She then adds a Nebula using the {\normalfont Sketch Mode + Region Mode} to draw a green mist on the canvas (\cref{fig:godzilla}~A1+A2), which the system repaints, blending it in with the rest of the scene.

For the second tile, she starts by generating a scene with overall description ``\textit{a cyberpunk city at night}''. She then wants to add a reusable {\normalfont neon billboard sign} asset. To do that, she adds a new node in the {\normalfont Tree View} (\cref{fig:treeview}), clears the canvas, and creates a ``\textit{neon billboard}''. Once happy with the result, she navigates back to the previous city scene using the {\normalfont Tree View} and drags and drops the neon sign into the scene. She masks out the borders of the new image asset, provides a short description of how the asset fits the scene, and the system blends the images.

Using the same techniques, she creates another tile showing a \textit{neon bridge at night} and a \textit{cyberpunk car at night}.  \participant{11}{} merges all four image tiles into one coherent image using the {\normalfont Global Tile View} (\cref{fig:global_tile_view}~A), adjusting the grid space and provides a short description that relates all tiles (\cref{fig:global_tile_view}C). She positions and resizes (\cref{fig:global_tile_view}B) the tiles to her liking and clicks {\normalfont Blend Tiles}. The system contextually fills in the space between the tiles (\cref{fig:global_tile_view}D, \cref{fig:example_blended_results}), and after some iterations.

\subsection{Technical Details}
Our prototype uses a client-server architecture. The client frontend is built with SvelteKit \cite{sveltekit} and Skeleton UI toolkit~\cite{skeletonui}, while the backend is built with FastAPI and runs on the python webserver uvicorn. We utilized the Stable Diffusion algorithm via HuggingFace's model hub \changenote{(512x512 pixels)}, which were run on a machine with 16GB GPU VRAM. To enable interactive painting, we used fabricJs\cite{fabricjs} and created two separate
canvases for sketching and region masks. When users initiated the diffusion process, we extracted inputs from the detail editor view. \changenote{For scene descriptions without additional input (e.g. sketches or regions) we generated an image from random noise based on the provided text input. If users provided an RGB sketch, we used it with added Gaussian noise to generate images that matches user's drawings following \citet{rombach2021highresolution}. Regions were transmitted as an array, with each entry containing a binary mask image and corresponding description. In our region-based painting feature (\cref{fig:region_drawing}), we extract multiple binary masks from a user-provided region segmentation, where white pixels correspond to the unique region color, and the remaining area is black.} Our region-based painting feature is inspired by \citet{balaji2023ediffi} \changenote{who proposed an approach allowing users to specify where elements should appear on the generated image. They combined a separate binary image mask, with dimensions matching the output image, with each word in an input text prompt. Notably, words in the user-provided text input exert variable influence on different parts of the image, with white pixels serving as indicators for a higher probability of an element appearing in the assigned segment.} We used an open-source implementation of this concept applied to Stable Diffusion\footnote{https://github.com/cloneofsimo/paint-with-words-sd}. \changenote{For blending image tiles, we obtained a binary mask with black pixels for image tiles and white pixels for \textit{empty} space. A Gaussian blur was applied to the mask, softening black tile edges for a smoother blend. The final image was generated by inputting this mask and user-created image tiles into the diffusion model. 
}

To track user interactions \changenote{(such as typing a scene description, drawing a sketch or a region, moving the tiles)}, we implemented a logging server on FastAPI, which utilized a Postgres database.

\section{EVALUATION}\label{section:main_study}
To evaluate the utility and use of \system{}, we conducted a user study with \mainN{} participants \changenote{which provide first insights into the following research questions:}

\begin{itemize}
    \item \RQ{1} How do users engage with generative AI for world building? 
    \item \RQ{2} To what extent does \system{} support the world building process?
\end{itemize}

\subsection{World Building Task}\label{section:task}
A set of prompts were created to inspire participants to build their fictional world. The prompts were intended to cover various types of visual world-building, such as fantasy maps or fictional landscapes. The prompts were open-ended and designed to allow participants to think creatively and explore a breadth of concepts (\cref{tab:design_prompts} in the Appendix).

We further encouraged them to explore not only maps but also other types of visual imagery depicting fictional worlds. While maps typically have specific characteristics such as a top-down view and a specific scale, fictional worlds can be composed of any objects that create a scene that does not exist in reality. We instructed participants to read the previous design prompts and focus on the composition of elements and the setting of the world (e.g., a fantasy world and everything such worlds entail). 
\subsection{Methods}\label{section:methods}
We conducted a first-use study online via Zoom and asked participants to think aloud during the user study to learn about their motivations and understand their potential challenges when interacting with the prototype. Our study involved a world-building task where participants used \system{} to create a fictional world (\cref{section:concept}). In addition to logging users' interactions, we conducted semi-structured interviews at the beginning and end of the world-building task to collect their open feedback about their prior experience and motivations for world-building and their overall experience working with \system{}. Finally, participants completed two questionnaires at the end of the task where they rated their experience with the different features of \system{}. For qualitative analysis, two researchers independently assigned inductive codes for a subset of the transcribed interviews. Then one researcher used these codes and notes taken during the interviews and thoroughly reviewed the full transcripts of all interviews to find further evidence for the thematic clusters identified in the previous step.

\subsection{Participants and Procedure}
We recruited the participants for the study over e-mail lists and personal contacts. All participants had experience building worlds (between 2 and 10 years). This included experience crafting DnD worlds, video game levels, creative writing, and landscape design. This study was approved through our institutional review process. 

The study consisted of four phases which we briefly outline below. Before the study, all participants completed a consent form and demographic questionnaire.

\textit{Pre-Interview ~ (10 minutes)} The first phase consisted of a short semi-structured interview with each participant. During this interview, we asked participants about their motivations and experience in building fictional worlds.

\textit{Tutorial ~ (10 minutes)} During the tutorial phase, participants were introduced to the prototype and given an overview of its features and functionalities. \changenote{One researcher explained the process of creating an example image tile using text, sketching or region painting. Meanwhile, the participants engaged with the tool by following the researcher's guiding instructions to become acquainted with the different ways of providing input.}

\textit{World Building Task ~ (60 minutes)} In the third phase, participants interacted with the software prototype via screen sharing and remote control. They were given a fictional world-building task to complete (\cref{section:task}). \changenote{We also asked them to briefly describe what they wanted to create} and we recorded their interactions with the software and resultant words.

\textit{Post-Interview and Questionnaire ~ (10 minutes)} After completing the world-building task, we conducted an interview with each participant to collect their feedback and impressions of the software prototype. In the questionnaire, we asked participants to rate the features of the prototype on a Likert scale. This questionnaire included questions about the various input modalities (text-only, text+region, text+sketching) and whether they think \system{} can speed up their regular approach to world-building. Finally, we administered a \textit{System Usability Scale} (SUS) questionnaire to evaluate the overall usability of \system{} which included questions about the complexity of the program, and related how easy it was to learn the program and to interact with it.

\subsection{Quantitative Findings}
Overall, we found participants leveraged all forms of interactions. %
In total, participants triggered the image generation process 229 times to generate scenes, maps, and assets for the individual image tiles, resulting in 2748 generated images. Additionally, users created 86 world compositions using the blending features.

In our first-use study, \mainN{} participants interacted with the prototype. However, some participants finished their world-building task before the official time ended, so they started a second session within the remaining time. Therefore, for the remainder of this section, we denoted each of the resulting 16 sessions with a participant id and the corresponding trial number ($\textit{P}_{X}~Trial~Y$).

\subsubsection{Relational vs. Quantifying Keywords}

We counted the number of words in 1) the \textit{text of each scene description} and 2) the \textit{text of each region description}.  We found that participants wrote differently for scene and region descriptions. Each participant created on average 2.05 regions with corresponding region descriptions per tile (Median=2, inter-quartile range 2). Region descriptions were typically shorter (Avg=4.2 words; Md=3; iqr=6) than scene descriptions (Avg=12.4 words; Md=11; iqr=7). Scene descriptions were primarily used to describe multiple objects and the overall style, while region descriptions were used to quantify or describe a specific object or element and to provide more localized instructions.

\begin{figure}
    \centering
    \includegraphics[width=\linewidth]{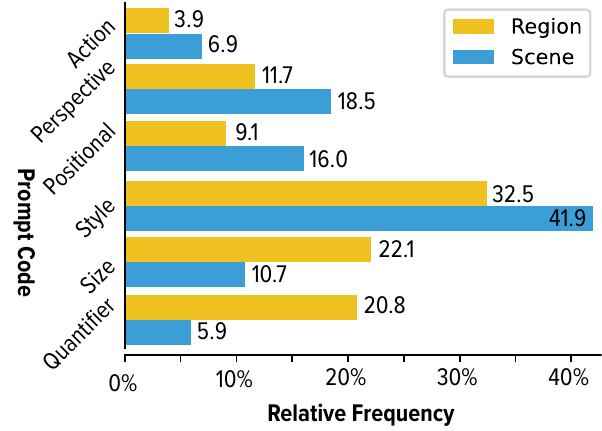}
    \caption{An overview of the codes extracted from the prompts for region descriptions and scene descriptions. See \cref{tab:codes} in the Appendix for an explanation of the codes. The blue and orange bars add up to one, respectively. Note that scene descriptions frequently contained style and perspective-related keywords, while region descriptions contained more size and quantifier keywords.
    }
    \Description{A chart showing the frequencies of the different codes extracted from user written prompts grouped by Region prompts and Scene Prompts.}
    \label{fig:prompt_style}
    \vspace{-1em}
\end{figure}

\textit{Position --}
We analyzed scene and region descriptions using a coding routine similar to that in \cref{section:methods}, using text prompts from user interaction logging data. The full list of codes can be found in \cref{tab:codes} in the Appendix. Our analysis found that scene descriptions contained more \textit{positional} keywords (e.g. \textit{surrounded by, above, north, south}) than region descriptions, which is consistent with participant observed behavior in the user study where they used the region drawing tool to specify spatial relations. This is shown in \cref{fig:prompt_style}.

\textit{Style --}
Scene descriptions also more frequently included \textit{style} (e.g. cartoon style, fantasy) and \textit{perspective} keywords (top-down view, isometric view). Participants used the UI in overall scene descriptions to define \textit{style} and \textit{perspective} keywords rather than repeating them for every region description. More generally, participants wanted to define these keywords once for the entire session and for all image tiles.

\textit{Action --}
Scene descriptions frequently included \textit{action} keywords, which relate one object to another (e.g. ``Mountain range \textit{running} north to south''). In contrast, region descriptions typically described only one object without trying to relate it to other elements in the world, which is done implicitly via drawing the regions.

\textit{Quantifier --}
Participants in our study used indefinite quantifiers (e.g. "a few trees", "many dirt roads") and size keywords (e.g. "a large", "a gigantic") more frequently in region descriptions to convey the number and size of elements in the world, instead of drawing separate regions for each.

\subsubsection{Composing Tiles}
\changenote{The final iteration with \system{} often involved rearranging and resizing the created image tiles (\cref{fig:example_final_generations_appendix}). The prompts at the global level were more generalized compared to the individual tile level prompts. They included keywords to remove visual artifacts such as border around the tiles during blending (\pquote{no [straight] edges, smooth collage}{2}{}, \pquote{seamless map}{1}{}) or introduce new objects to connect the tiles (\pquote{[...] islands connected by bridges}{5}{}, \pquote{[...] an ocean in the center}{8}{}). Although, keywords related to style and perspective were also used, global keywords didn't affect already created image tiles since \system{} does not support perspective and style matching across tiles post image tile creation.
}

\subsubsection{Interaction traces}
\begin{figure}
    \centering
\includegraphics[width=\linewidth]{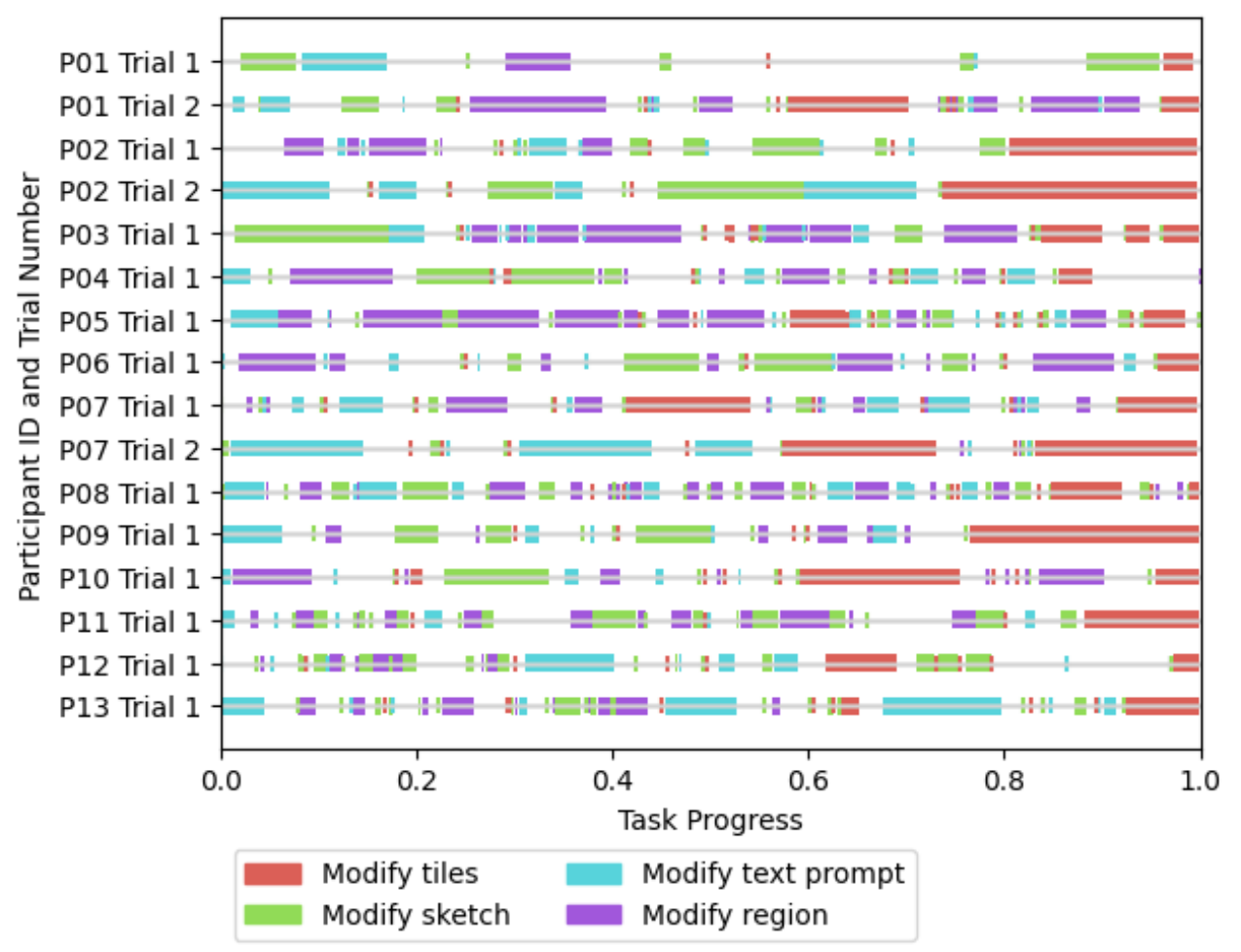}
    \caption{The graph shows the distribution of participants' interaction traces while they worked on the interactive prototype. Three participants completed the task early and started a second world-building session, resulting in a total of 16 sessions instead of 13. Note that most participants started with a text prompt when first interacting with \system{}. }
    \Description{A graph with multiple lines. Each line shows the normalized time users have spend engaging in different activities such as modifying tiles, modifying the sketch, etc.}
    \label{fig:interactiontraces}
    \vspace{-1em}
\end{figure}

Users' editing behavior was analyzed by filtering four representative actions from the interaction logging data (Figure \ref{fig:interactiontraces}). These actions include modifying tiles, sketches, regions, and text prompts. Examples of modifying tiles include repositioning and scaling them in the \textit{Global Tile View} (\cref{fig:teaser} (A)) while modifying sketches involves drawing on the Sketch Canvas. Modifying regions involves adding, drawing, and describing new regions, and modifying text prompts only involves editing the scene description in the \detailview{}.

\textit{Bootstrapping the World with Text --} \label{section:bootstrap}
Participants in 12 out of 16 sessions began creating their world with a text-only description, likely due to the fast and lightweight nature of the method, as reported by participants in the  study: \pquote{I do like just having the general big description, and just seeing what it comes up with.}{11}{}. Participants in the other 4 sessions started with an initial sketch of their worlds. These participants already had an image with a rough structure in mind. Using the sketching tools directly, they could sketch out their mental image. %

\begin{figure}
    \centering
    \includegraphics[width=\linewidth]{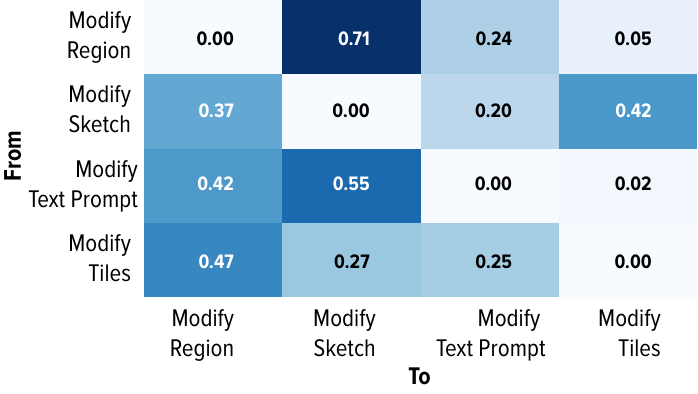}
    \caption{An overview of the action transition frequencies aggregates across all sessions. The number in each matrix cell represents the relative transition ratio at which a participant transitioned from action Y (rows) to action X (columns).}
    \Description{A matrix that highlights activity transition probabilities.}
    \label{fig:transition_matrix}
\end{figure}

\textit{Moving from Coarse to Detail --}
We computed and aggregated the transition matrices between creation operations over all participants and trials to analyze editing behavior (\cref{fig:transition_matrix}). Overall, participants have transitioned between all available edit operations. However, modifying regions frequently preceded the blending process the last action in the world-building process. This observation, together with our previous observations (\cref{section:bootstrap}) suggest that participants transitioned from coarse actions, such as text prompts, to fine-grained editings, such as sketching and region painting, to add details to their image tiles. This was consistent with the previous observation that participants used text prompts to quickly bootstrap the world-building process.

\textit{Summary --} \changenote{The quantitative results indicate that users engaged with \system{} in different ways \RQ{1}. They generally moved from coarse to detail, i.e. using text and regions to populate the individual tiles before sketching in the details. On the global composition level, participants explored alternative worlds by revising both their prompts as well as tile compositions.} Differences in language used in their prompts imply that users perceive scene and region prompts differently and convey input information implicitly through both text and sketching.

\begin{figure}
    \centering
    \includegraphics[width=\linewidth]{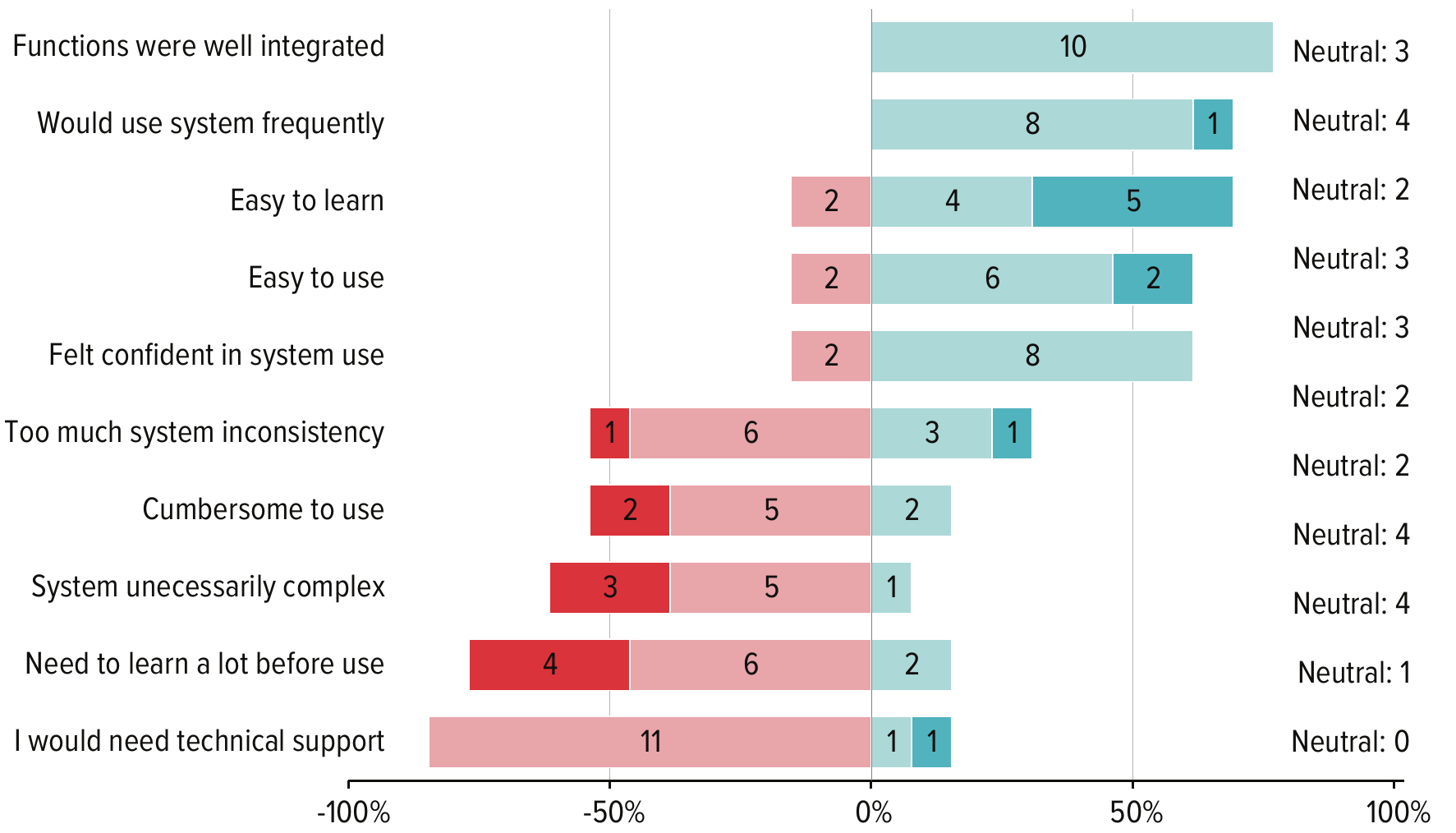}
    \caption{An overview of the responses for the System Usability Scale questionnaire.}
    \Description{A chart showing user responses to the SUS questionnaire.}
    \label{fig:sus}
    \vspace{-1em}
\end{figure}

\subsection{Design Insights and Improvements}\label{sec:results_design}
Overall, the participants felt that the tool was easy to use and helped them perform the world-building task (\cref{fig:sus}). Our observations and participants' comments during the interview revealed that some system features required further iteration.

\textit{Related Keywords --}
As noted by \citet{liu2021design} and \citet{liu20223dalle}, developing relevant keywords for text-to-image generations is challenging. While participants in the study were primarily concerned with building their fictional world, they felt burdened to think of keywords that are relevant to their world, such as stylistic keywords and perspective. They wished the system would automatically insert those keywords and match the style across all tiles.

\textit{Direct Manipulation --}
There was a desire to more directly edit the result image, with one participant explaining \pquote{It would have been nice to be able to manipulate elements in the generated image directly. For example, after I created that tile with a crossroad, I wanted to select that road as a spline object that could be directly manipulated.}{2}{main}. This would allow for greater control and precision in the image creation process, though comes with a large set of technical challenges.

\begin{figure}
    \centering
    \includegraphics[width=\linewidth]{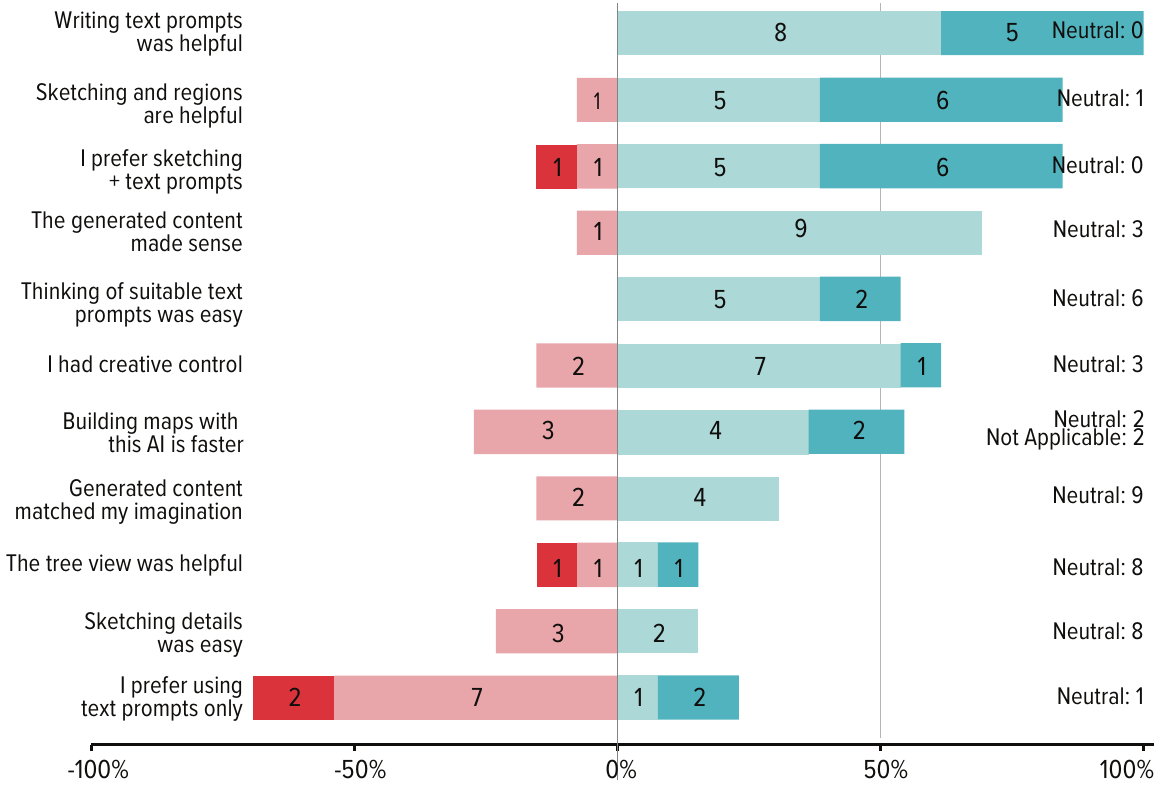}
    \caption{An overview of the responses from the  final questionnaire.}
    \Description{A chart that shows user responses to the final questionnaire.}
    \label{fig:final_questionnaire}
    \vspace{-1em}
\end{figure}

\subsection{Qualitative Findings}

\subsubsection{Working with Region-Based Descriptions}\label{section:region_based_interaction}
We observed two distinct strategies when participants interacted with the region-based descriptions.

\textit{Narrative first, then drawing --}
When defining regions, some participants created the text segments first and typed the full description of each region before deciding how these regions are composed on the canvas. This is in line with findings from the formative study (\cref{section:formative_results}) on starting with a concept of the fictional world first before committing to a specific instantiation of that world.

\textit{Drawing first, then narrative --}
On the other hand, some participants first thought about the elements' overall composition before writing down a detailed description for each element. When asked whether they had a concrete image in mind, they replied with: \pquote{I don't have anything specific in mind, but rather a rough idea of the elements I want in that picture.}{13}{}

\subsubsection{Users preferred Multi-Modal Input}
While all participants agreed that writing text prompts to generate images was helpful, the majority (11 out of 13) preferred multi-modal input over using text-prompts only (\cref{fig:final_questionnaire}). Deriving suitable text prompts for image generation was not considered difficult, however, we observed during the study that participants valued content that described the core-elements of their fictional-world (e.g. a magic forest, a mage rabbit), more than keywords which related to overall rendering features such as stylistic and perspective keywords. Participants also responded that they preferred adding sketches and painting in addition to their text input. From our observations we note that some participants (\participant{2}{}, \participant{3}{}) found it challenging to sketch-in details with the remote controlled computer using the mouse, while other participants saw this technical constraint as a strong motivator for generating images from rough sketches without the need for precise 2D input (\participant{6}{}) using \system{}.

\subsubsection{Building World by Parts}
During the formative study, we found that not all parts of an image hold equal importance for the creator. This is particularly evident in Dungeons and Dragons (DnD) games, where certain areas of interest contain more detail and are the sites of important events. Observations from our first-use study underpinned the previous finding by showing that participants invested significant effort in creating highly detailed individual tiles, and appreciated the ability to focus on these tiles while leaving the empty spaces between them to be filled in automatically by \system{} to smoothly combine the tiles. Additionally, one participant noted that \pquote{[t]he tiles have lots of detail which naturally draws the viewers eye to those points of interest.}{1}{}. Another participant found blending to be useful to explore different world compositions quickly:
\pquote{
 You know, sometimes [...] I know that I want there to be like a lake over here in a city over here. But I don't really care what else is in there [...]. Let's throw together some interesting stuff and see how it blends together, and then start adding on from here.}{4}{}.

\subsubsection{Materialise Stream of Consciousness}\label{sec:materialize}
Participants commented that they sometimes struggled to track their thoughts because \pquote{ideas flash up and vanish}{13}{} before they had the chance to fully develop that idea. However, using image generation has aided them in quickly capturing their train of thought. One participant further noted \pquote{I'm one of those people who has a very blind inner eye, so I can't visualize things in my head. I have to put it in front of me in order to make any sense of it visually.}{11}{}. Using the multi-modal image generation system has enabled them to refine their initially \textit{vague ideas} by allowing them to continuously add details to their creations. Here, one participant noted that \pquote{seeing how all seamlessly blend, I now had a clearer vision of how I want to compose the elements in the world}{3}{}. 

\subsubsection{Perception of control over the AI}
Participants were generally conscious that they were interacting with an AI system. During the world-building task, we frequently observed participants questioning whether the system understood their commands or intentions. As a result, participants appeared to be more understanding when \system{} failed to generate exactly what they had in minds. In such instances, participants began to consider alternative ways of expressing their intent. They sometimes reformulated their text input to the system or included supplementary information for the system, such as drawing a sketch. For example, one participant wanted to create a nebula on a night sky using the region-based painting tool, but the system did not produce the desired result. In response, the participant decided to add more information to the input by also adding a sketch in addition to the region-based painting (see \cref{fig:godzilla}). \changenote{However, some participants were discouraged by their initial failed attempts to author their envisioned image tile, feeling that the system was inconsistent (\cref{fig:sus}) when generating new images (\participant{4}{}, \participant{7}{}, \participant{8}{}). For example, while \participant{4}{} could use \system{} to generate \textit{``ducks in a pond''} or \textit{``a house surrounded by a forest''}, he was struggling to generate a complex scene such as \textit{``a top-down view of a mountain rift running North to South''} and wanted to insert a \textit{``a fantasy art of a mouth of a mine with mine carts arranged around''} and a \textit{``top down view of a yard surrounded by tents and roman soldiers''}. Creating such a scene would have required more time to create relevant assets such as the mine or the war camp. 
}

\subsubsection{Feedback on the Tree View}
Overall, we found that those participants who used the tree view explored this concept to create new image assets and blend them back into their scenes. One participant mainly used this feature to introspect his past interactions. The tree view enabled him to be more confident in exploring alternative generations because it offered a way to revert to the original image. \pquote{I really liked the tree view, because I use a lot of [...] platforms and I'm often hesitant to continue iterating on an idea [...] because it's hard to get back to the original image. Sometimes it kind of gets lost, even if it's somewhere in the history it's up to you to find the original image that you originated from [...]}{10}{}. Another participant commented that she wanted to create \pquote{unlimited nodes and add them directly as image tiles to the global canvas}{12}{}. 
However, the responses in the final questionnaire (\cref{fig:final_questionnaire}) indicated that participants were generally neutral about the \textit{Tree View} in this study. We reflect on this observation in the discussion.

\subsubsection{Feedback on the Generated Worlds}
\begin{figure}
    \centering
    \includegraphics[width=\linewidth]{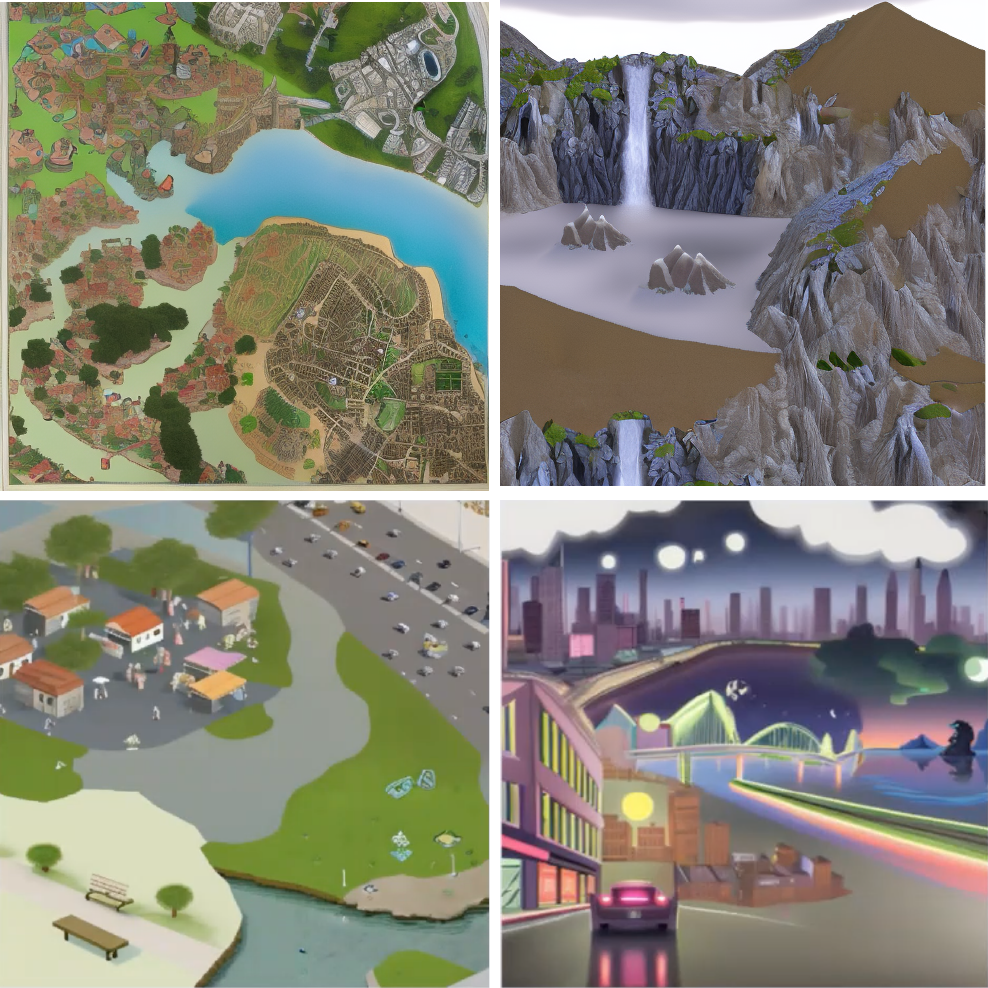}
    \caption{Figure shows multiple worlds that users have built using \system{}. Some show a fictional map, while others depict a scene in a fictional world.}
    \Description{A collage showing some generated images depicting participants envisioned worlds.}
    \label{fig:example_blended_results}
    \vspace{-1em}
\end{figure}

Participants in our study have found that the generated worlds made sense, although they did not always match their prior imagination. This included the generated individual tiles as well as the overall blended tiles (\cref{fig:example_blended_results})

During the interview 11 out of 13 participants commented positively about the blended results, highlighting that the system was capable of sensibly filling in the gaps between the created image tiles. The other two participants generally liked the idea of blending different parts of their image but found it challenging to create a seamless blend of all their tiles.

\subsubsection{Comparisons against world-building without Generative AI}
\changenote{
When participants reflected on their experience with \system{}, two participants commended the fast generation of an initial map draft by focusing only on a few \textit{key areas} (\participant{1}{},\participant{9}{}). Nevertheless, full utility of the generated maps required better style and perspective alignment across all tiles (\participant{6}{}, \participant{9}{}, \participant{8}{}). For example, \participant{8}{} in found it difficult to align a tile with \textit{``a cartographic rendering of an arctic tundra''} with a \textit{``cartoon style desert island with war camps''}.\\
Aside from DnD maps, one participant particularly liked the ability to customize tiles. He compared the tile generation process using \system{} and his prior experience with in-game map builders noting: \pquote{In the past, while working with tile-based map editors, I frequently found myself searching for compatible tiles, for instance, connecting streets. However, with this system, I would simply provide sufficient space between the tiles and let the system determine the best way to merge them}{6}{}. \participant{11}{} in particular found that \system{}'s proposed workflow closely matches her own world-building process: \textit{``I do the more fine details and then go. Oh, wait! I should do something [...] a little bit more broad. So for me [the workflow] wasn't anything new. It was kind of a more natural flow for me.''}
}

\changenote{\textit{Summary --} In relation to \RQ{2}, our findings demonstrate that the suggested workflow of \system{} effectively complemented participants' existing world building process, enabling them to swiftly produce a rendition of their envisioned worlds. During a 45-minute user study, each participant successfully generated a version of their world. While \system{} facilitated the rapid creation of an initial draft, participants also acknowledged the need to address appropriate scaling, stylization, and perspective coherence among the already generated image tiles in order to accurately represent their envisioned world. }

\section{Discussion}\label{section:discussion}

\subsection{Speed vs. Quality Trade-Off}

We observed that the quality of the blended results varied among participants. Some participants could seamlessly blend multiple tiles, while others required multiple iterations. We observed three main factors that influenced the quality of the blended result in our study: 1) Tiles with similar styles were easier to blend. 2) Participants had to consider the logical structure of tile composition, such as creating a single horizon across multiple tiles to achieve a seamless blend. Ambiguous positioning of the horizon could result in a less seamless blend. 3) Tile complexity, including the length of the text prompt, number of regions, and style parameters, also affected blending quality. More detailed tiles were harder to blend seamlessly.

We experimented with MultiDiffusion \cite{bar2023multidiffusion} and found it produced better results than stable diffusion but was significantly slower (over 30 seconds for a 768x768 pixel image). Therefore, we prioritized speed and used stable diffusion for image generation to allow users to iterate quickly. Our main objective was to analyze how users developed their fictional world. Some participants were satisfied if the approximate content of their initially created tiles was preserved for the final blending. They were also willing to wait for the final rendering once they were satisfied with the overall composition. To address this, future work could include a slower, quality-preserving mechanism like MultiDiffusion to be used as a post-processing step for the final rendering.

\subsection{Creative Use of Model Bias}

We have found that stable diffusion \cite{rombach2021highresolution} excels in generating images where the elements fit naturally into the scene, such as a spider in a forest, but often faces challenges when generating images of concepts that do not belong together using only text prompts, such as a snow-covered mountain ridge on a tropical island. It has been found that LLMs, sometimes "hallucinate" facts and struggle to generate coherent storylines \cite{RoberGPT3WGF2021}. However, our users' feedback indicated that visual hallucinations \cite{Berov2016VisualHF} can be a desirable attribute that allows users to create unconventional worlds. Therefore, our prototype offers two possible interactions to combine different concepts. The first solution involves merging tiles that portray disparate concepts, while the second allows users to blend arbitrary images into the scene using a region mask. 

Related work has improved the synthesis of images with disparate concepts  \cite{Gafni2022MakeASceneST, balaji2023ediffi}. With \system{} we complement these systems by providing a tool that enables users to interactively define blending on three levels: 1) blending using text-only descriptions, 2) blending by also providing sketching and region-masks for disparate concepts, and 3) blending of multiple image tiles.

\subsection{Limitations and Reflections on Methodology}
Our study was conducted with a pool of participants who shared a common interest and proficiency in technical systems. Many had prior exposure to generative image systems, which allowed them to engage easily with the system during the study. We note that participants without prior experience in image editing may require additional time to familiarize themselves with the system. This learning curve may vary depending on the individual's technical knowledge and experience level. \changenote{The user study suggested that participants mainly required help to find domain-specific language related to world-building, e.g. perspective, style. Nonetheless, we ensured that all participants received guidance and suggested relevant keywords based on the verbal descriptions of what they wanted to create}. \changenote{Future systems can include a LLM to suggest relevant keywords to lower the threshold for using the system \cite{liu20223dalle}}.

During the study, we found that participants were focusing more on building the individual image tiles, which occupied most of their time. Given the time limit of the user study, few participants interacted extensively with the \textit{Tree View}. However, those who did comment positively about it. We believe that such a tool is best explored over a longer period and over distinct sessions to enable users to introspect their past behavior and explore a wider variety of scenes as they create their world.

\changenote{In this study we focused on the multi-modal input techniques to leverage a generative AI for world-building. However, we did not consider lore, musical score or character building which are also an essential component of successful worlds.}

Our study identified certain limitations with respect to the quality and speed of \system{}. We found that the generated images sometimes did not accurately capture all the concepts mentioned in users' text descriptions. Additionally, users had to wait for 6-8 seconds before the first batch of images was generated. While we expect that the development of pre-trained generative AI models will continue to improve the accuracy and speed of image generation, it is important to acknowledge that human language remains inherently ambiguous. As such, we need to explore additional methods to help users express their creative vision when working with generative AI models.

\subsection{Expressive Prompting}

Current user interfaces for prompt-based models \cite{midjourney2023, Ramesh2022HierarchicalTI} tend to promote a one-shot image generation approach, wherein users can only modify the text to influence the generation process. However, our user study highlighted a key insight: participants tend to create their world models in parts, using sketching and text inputs to communicate positional information. To support this design process, \system{} offers users the ability to sketch and paint regions in addition to entering text prompts. Furthermore, \system{} incorporates hierarchical support by separating tile blending from tile creation.

Based on our improved understanding of how users integrate multi-modal input in the process of building worlds, we have identified two distinct dimensions that demonstrate how \system{} enables more expressive prompting beyond the limitations of the "click-once" interaction paradigm.

\textit{Hierarchical Prompting --}
\system{} facilitates hierarchical prompting by enabling users to define prompts across three levels. Firstly, at the Image Tile Canvas level, users can provide text prompts to create a base scene, thus setting the stage for their world-building process. Secondly, at the Sketch and Regions level, users can add text prompts to sketches and regions to provide structural guidance to the system. Finally, users can add text prompts to blend multiple tiles together at the Global Tiles level.

\textit{Spatial Prompting --}
In addition to its other features, \system{} also allows participants in our study to modify prompts spatially through non-textual interactions. This capability is facilitated through two methods: firstly, users can convey spatial prompt information through sketching, thus engaging in what we term "Prompting through Painting." Secondly, users can move image tiles to convey prompt information, which we call "Prompting through Dragging."

In summary, \system{} introduces two distinct dimensions that enhance users' ability to interact more expressively with prompt-based models. \changenote{However, we believe that expressive prompting is not limited to world-building, but can also aid the composition of any complex image. Traditional photo editing includes a layering system to allow users to organize and structure their images manually. Spatial and hierarchical prompting augments this interaction and allows participants to quickly explore blended image compositions.}

\section{Conclusion}
In summary, we have looked beyond the "click-once" image generation interaction paradigm for prompting generative AI, and we discovered through a first-use study with \system{} that users leveraged all inputs (text, sketch, and region masks) in combination. Crucially, participants expressed their creative vision not only through textual descriptions but also through non-textual interactions with the system. Based on our findings, we propose two expressive prompting concepts as part of \system{}'s graphical UI, supporting: 1) \textit{hierarchical prompting}, which involves the use of layered prompts, and 2) \textit{spatial prompting}, which allows users to spatially arrange prompts. With \system{}, we illustrate how these prompting concepts aid the fictional world-building process. Beyond this use case, we see expressive prompting as a general concept to inspire user interfaces that support users' complex workflows with prompt-based AI models.

\begin{acks}
We thank Justin Maltejka, Jo Vermeulen, Bon Adriel Aseniero, David Ledo, Qian Zhou and Daniel Buschek for providing valuable feedback on this work. 
\end{acks}

\bibliographystyle{ACM-Reference-Format}
\bibliography{main}

\appendix

\setcounter{figure}{0}
\renewcommand\thefigure{\thesection.\arabic{figure}}
\setcounter{table}{0}
\renewcommand{\thetable}{A.\arabic{table}}
\section{Appendix}\label{sec:appendix}

\begin{table*}
\centering
\footnotesize
\newcolumntype{L}{>{\raggedright\arraybackslash}X}
\renewcommand{\arraystretch}{1.4}
    \begin{tabularx}{\textwidth}{lL}
    \multicolumn{1}{l}{\textbf{Topic}} 
        & \multicolumn{1}{l}{\textbf{Design Prompt}} \\
    \midrule
    Fantasy World        
        & Imagine a world unlike any other, where the terrain is so varied and unique that each step you take leads you into a new adventure. The air is thick with mystery and magic, and the landscapes range from towering mountains to sprawling forests, vast deserts to shimmering oceans. Now, as a cartographer, it's your job to bring this world to life with your maps. With each stroke of your pen, you have the power to transport your readers to this magical realm and inspire them to explore every inch of its diverse and detailed landscapes. Are you ready to create a map that will take your readers on the adventure of a lifetime? \\
    Landscape architect   
        & As a landscape architect, your task is to create a fictional world that is a realistic simulation of the real world. The world should be designed with the same level of detail and accuracy as a real-life landscape, incorporating features such as realistic elevation changes, accurate water flow patterns, and vegetation that is appropriate to the climate and terrain  \\
    Computer Game
        & Create a computer game map that is both visually stunning and functionally complex, offering players a dynamic and engaging environment that requires strategic thinking and quick reflexes to outsmart opponents. Incorporate varied terrain, structures to capture and defend, hidden paths and secret locations, and balance gameplay mechanics to ensure a fair yet challenging experience. Transport players to a world of intense strategy and high stakes, challenging them to work together as a team to overcome their opponents in a game map that will be engaging, immersive, and memorable.\\
    Treasure Hunting    
        & Embark on an exciting adventure as a treasure hunter, exploring a map that is full of hidden riches and ancient artifacts waiting to be discovered. Use intricate puzzles, traps, and obstacles to create a challenging and engaging experience that will keep players on their toes. Populate the map with unique and exotic locations such as forgotten tombs, lost cities, and hidden underground chambers. Create a visually stunning landscape with diverse terrain, beautiful vistas, and atmospheric lighting that immerses players in a thrilling world of discovery and adventure. \\
        \bottomrule
    \end{tabularx}%
\vspace{1em}
\caption{An overview of the design prompts.}
\Description{A table listing all world-building prompt for the user study.}
\label{tab:design_prompts}
\end{table*}

\vspace{-1em}

\begin{table*}[h]
\centering
\footnotesize
\newcolumntype{L}{>{\raggedright\arraybackslash}X}
\renewcommand{\arraystretch}{1.4}
    \begin{tabularx}{\textwidth}{llL}
    \multicolumn{1}{l}{\textbf{Code}} 
        & \multicolumn{1}{l}
         {\textbf{Description}}
        & \multicolumn{1}{l}
        {\textbf{Examples}} \\
    \midrule
    Size 
        & Indicates relative sizes of the world's elements.
        & large; small; high; giant; tall; tiny; big \\ 
    Positional 
        & Terms that express how multiple objects are positioned in the image.
        & surround; above; side by side; around; underneath; middle; on both sides; bottom; left; corner; north; south; next to; in the carribbean; contain; in Rome; between; split by; inland; west; east \\ 
    Action 
        & Describes whether objects's actions.
        & hunting; selling; erupting; on fire; sits; running; coming; reach; wear; explosion; smoking; painting; holding; extending \\ 
    Quantifier 
        & Relates to the quantitiy of an object
        & many; few; dense; some; a lot of; lots of; singular; four; two; several \\ 
    Style 
        & Expresses the style of the image.
        & concept art; map; anime; cyberpunk; 1950; antique; cartoon; japanese; medieval; fantasy; futuristic; cartographic; geographical \\ 
    Perspective 
        & Relates to the viewpoint or perspective of the image.
        & 2d; top down; horizontal; skyline view; view; isometric; bird \\
    \bottomrule
    \end{tabularx}%
    \vspace{1em}
\caption{An overview of codes developed for the analysis of scene and region descriptions.}
\Description{A table with three columns: Code, Description and Examples. The table lists all codes extracted from user written text prompts.}
\label{tab:codes}
\end{table*}

\vspace{-1em}

\begin{figure*}
    \centering
    \includegraphics[width=0.9\linewidth]{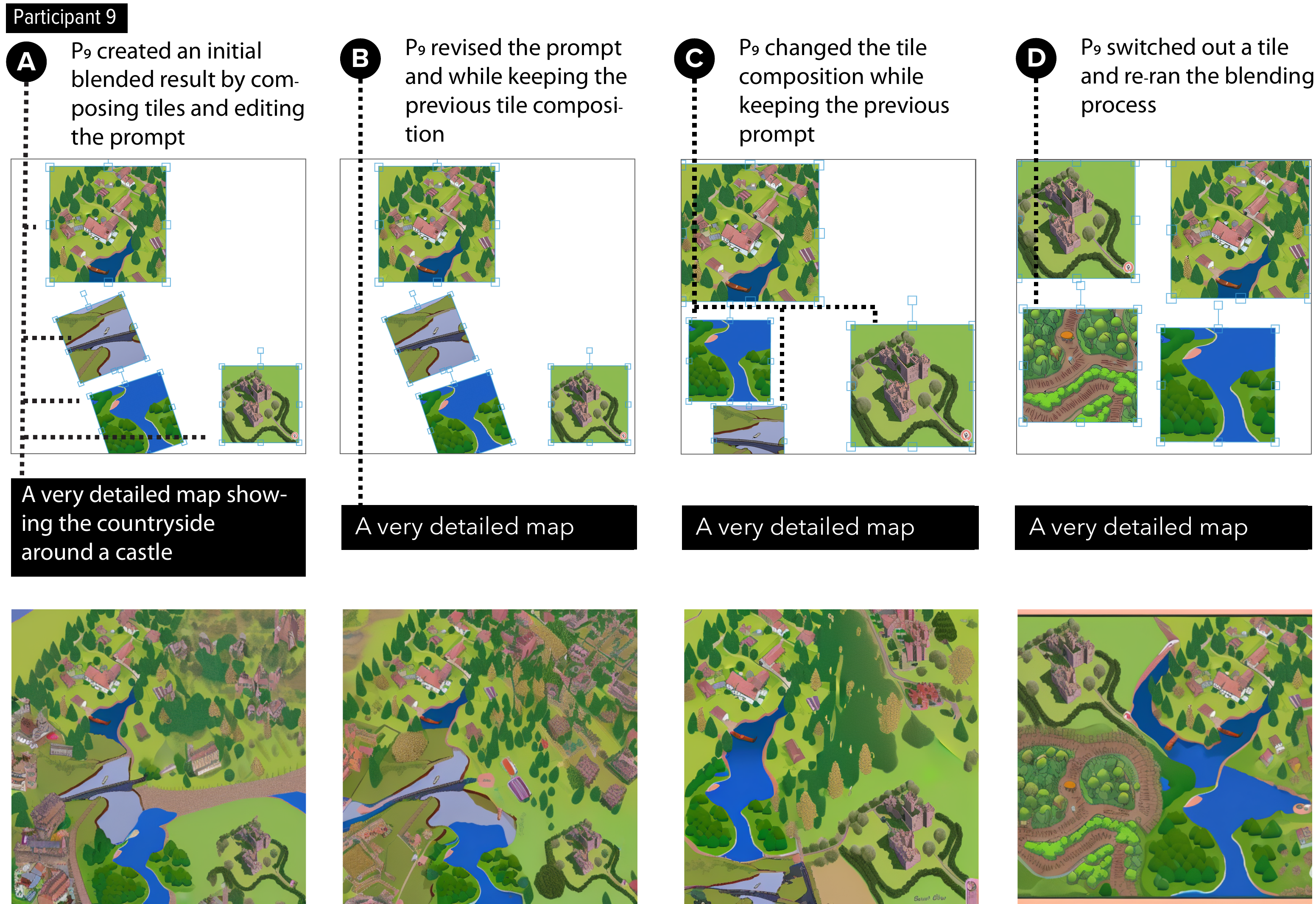}
    \caption{Multiple world blending results with \system{} (\participant{9}{}). (The top row) shows the four input tiles. (The bottom row) is the final blended image depicting participants fictional worlds. \participant{9}{} designed a top-down of a map consisting of a castle, a small village, farmlands, and a large river. \\}
    \Description{An overview showing how Participant 9 iterated in the global tile view once all tiles have been created.}
    \label{fig:example_final_generations_appendix}
\end{figure*}

\end{document}